\documentclass[acmtog,authorversion]{acmart}
\usepackage{wrapfig}
\usepackage{multirow}

\usepackage[export]{adjustbox}

\AtBeginDocument{%
  }

\setcopyright{acmlicensed}
\copyrightyear{2026}

\acmYear{2026}
\acmVolume{45} 
\acmNumber{4} 

\acmConference[SIGGRAPH '26 Conference Papers]
  {Special Interest Group on Computer Graphics and Interactive Techniques Conference Conference Papers}
  {July 19--23, 2026}{Los Angeles, CA, USA}

\acmJournal{TOG}
\acmArticle{164}

\definecolor{AAA}{rgb}{1.0, 0.13, 0.32}
\definecolor{BBB}{rgb}{0.2, 0.1, 1}
\definecolor{CCC}{rgb}{0.0, 1, 0}
\definecolor{DDD}{rgb}{0.0, 0, 1}

\usepackage{xcolor}
\usepackage{colortbl}
\usepackage{graphicx} 

\definecolor{headerblue}{rgb}{0.8,0.87,0.94}
\definecolor{rowgray}{rgb}{0.95,0.95,0.95}

\definecolor{myorange}{HTML}{ff9a33}

\definecolor{onec}{HTML}{fe218b}
\definecolor{twoc}{HTML}{fed700}
\definecolor{tric}{HTML}{21b0fe}

\usepackage[linesnumbered,ruled]{algorithm2e}

\begin{document}

\title{Structural MAT: Clean and Scalable Medial Axis Simplification via Explicit Surface Correspondence}

\author{Pengfei Wang}
\affiliation{%
  \institution{Shandong University}
  \city{Qingdao}
  \country{China}}
\email{pengfei1998@foxmail.com}
\orcid{0000-0002-2079-275X}

\author{Shuangmin Chen}
\affiliation{%
  \department{School of Information and Technology}
  \institution{Qingdao University of Science and Technology}
  \city{Qingdao}
  \country{China}}
\affiliation{%
  \institution{Shandong Key Laboratory of Deep Sea Equipment Intelligent Networking}
  \city{Qingdao}
  \country{China}}
\email{csmqq@163.com}

\author{Dongming Yan}
\affiliation{%
  \institution{Institute of Automation, Chinese Academy of Sciences}
  \city{Beijing}
  \country{China}}
\email{yandongming@gmail.com}

\author{Ying He}
\affiliation{%
  \institution{Nanyang Technological University}
  \city{Singapore}
  \country{Singapore}}
\email{yhe@ntu.edu.sg}

\author{Shiqing Xin}
\authornote{Corresponding authors.}
\affiliation{%
  \institution{Shandong University}
  \city{Qingdao}
  \country{China}}
\email{xinshiqing@sdu.edu.cn}

\author{Changhe Tu}
\authornotemark[1]
\affiliation{%
  \institution{Shandong University}
  \city{Qingdao}
  \country{China}}
\email{chtu@sdu.edu.cn}

\author{Wenping Wang}
\affiliation{%
  \institution{Texas A\&M University}
  \city{College Station}
  \country{United States of America}}
\email{wenping@tamu.edu}

\renewcommand{\shortauthors}{Wang et al.}

\acmSubmissionID{182}
\begin{abstract}
The Medial Axis Transform (MAT) is a complete shape descriptor capable of reconstructing the geometry of the original domain. A high-quality MAT should not only facilitate high-fidelity reconstruction but also capture structural features---for instance, by aligning the MAT boundary with the locus of rolling ball centers within fillet regions. However, computing such an ideal MAT remains a significant challenge, particularly when the input is a discrete triangle mesh. 

In this paper, we follow the established technical pipeline of initializing the MAT via a 3D Voronoi diagram of surface samples and subsequently simplifying the Voronoi structure through a QEM-like scheme. Our key insight is to explicitly track the correspondence between MAT vertices and surface regions throughout the progressive simplification process, ensuring that the resulting MAT triangles accurately reflect the intrinsic symmetries between surface patches. We translate these geometric requirements into a suite of priority control strategies that govern the sequencing of edge collapses.

Through extensive evaluation against state-of-the-art MAT algorithms, we validate the strong performance of our approach regarding runtime efficiency, structural alignment, boundary regularity, triangle quality, and robustness to noise.
Our resulting MATs remain highly expressive for both articulated shapes and CAD models, even under extreme simplification---effectively capturing the global structure of complex geometries with only a few hundred vertices.
Finally, we showcase the utility of our approach through two potential applications: capturing the locus of rolling ball centers within fillet regions, a structural capability not previously demonstrated in the existing literature, and surface extraction from unsigned distance fields, where the medial axis of the $\epsilon$-isosurface naturally yields a clean single-layer result.

Source code is available at \url{https://github.com/sssomeone/structural-mat}.
\end{abstract}



\begin{teaserfigure}
  \includegraphics[width=\textwidth]{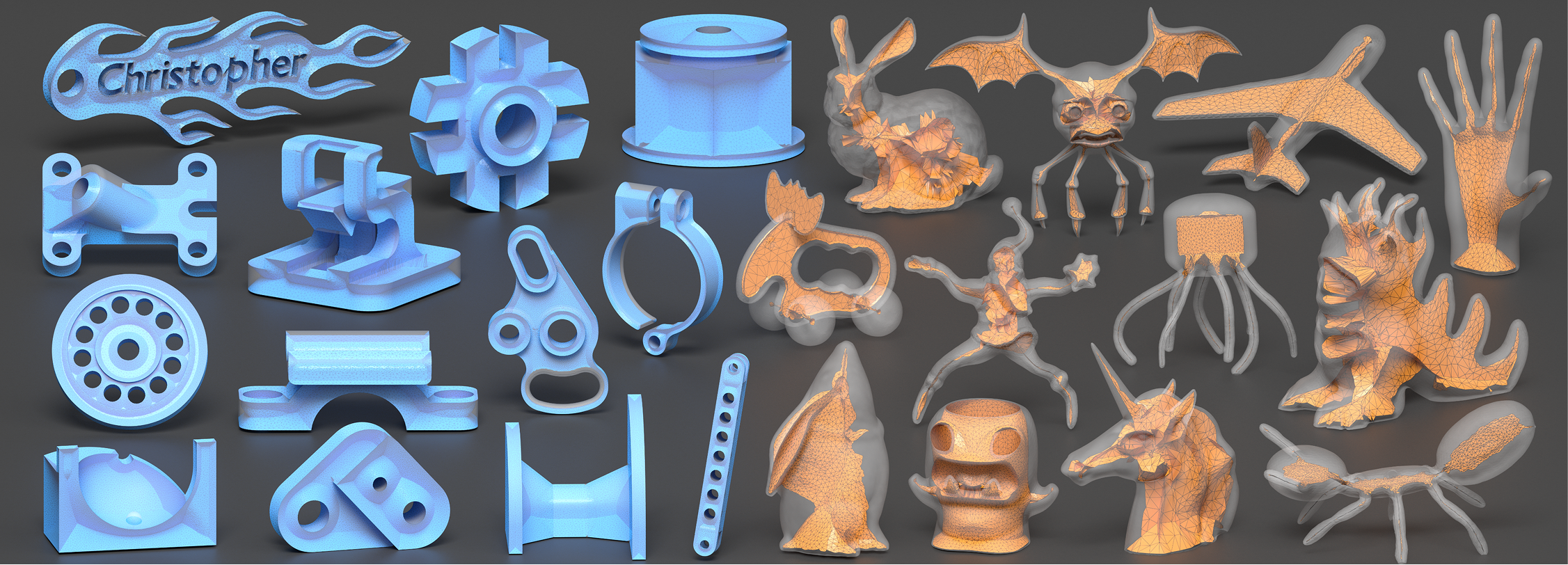}
  \caption{Representative medial axis results computed by our method on CAD models with sharp features and organic models with smooth surfaces. The surface-guided simplification framework ensures clean medial boundaries, accurate feature alignment, and high mesh quality across diverse geometric inputs.}
  \label{fig:teaser}
\end{teaserfigure}

\maketitle

\section{Introduction}

Mathematically, the medial axis of a bounded shape is defined as the set of points with at least two nearest points on the boundary---or equivalently, as the locus of centers of maximally inscribed spheres. Together with their associated radii, these points form the Medial Axis Transform (MAT), a compact and complete descriptor that encodes a shape's topological structure and local thickness. Due to its powerful geometric abstraction capabilities, the MAT is a cornerstone of shape understanding, supporting a wide range of downstream applications including shape analysis~\cite{dou2020top,fu2022easyvrmodeling,hu2019mat,noma2024surface}, shape decomposition~\cite{zhou2015generalized,lin2020seg}, pose analysis~\cite{yang2021learning,dou2023tore}, and animation~\cite{lan2021medial,lan2020medial}.

The quality of a medial mesh representation is typically evaluated by two standards: its ability to accurately reconstruct the original surface and its capacity to effectively inherit structural features.
Furthermore, the triangle quality of the MAT mesh itself is a critical factor for downstream stability. Most prior research has prioritized reconstruction accuracy while paying less attention to structural alignment. For instance, Q-MAT~\cite{10.1145/2753755} employs progressive edge collapses where the collapse priority is based primarily on the minimization of reconstruction loss. However, as simplification proceeds, the connection between a MAT vertex and the underlying surface symmetry weakens, leading to structural drift. 

\begin{figure}[t]
  \centering
  \includegraphics[width=0.99\linewidth]{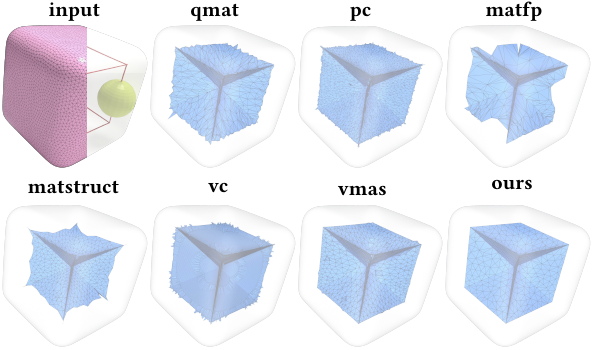}
  \caption{Medial axis computation on a filleted cube model. The true medial axis boundaries should align with the centerlines traced by the rolling sphere centers in the filleted regions. Our approach captures this structure, demonstrating its capability to faithfully sense and represent intrinsic structural features.
 Notably, VMAS~\cite{10.1145/3680528.3687678} is evaluated with the correction step disabled due to optimization instability.}
  \label{fig:roundcube}
\end{figure}

While recent works such as MATFP~\cite{wang2022matfp}, MATTopo~\cite{wang2024mattopo}, and MATStruct~\cite{10.1145/3757377.3763840} have improved geometric fidelity and topological correctness for CAD models through feature-aware optimization, challenges remain in balancing computational efficiency, mesh quality, and geometric accuracy. Despite these considerations, existing methods struggle to produce regularized, zigzag-free boundaries. Figure~\ref{fig:roundcube} illustrates a toy model with round fillets; it is evident that previous methods fail to align the MAT boundary with the locus of rolling ball centers within fillet regions.

In this paper, we shift the focus toward the structural symmetry of the original surface to produce a structurally reliable MAT. We adopt the established technical pipeline of initializing the MAT via the Voronoi diagram of dense surface samples followed by progressive simplification via edge collapses. Our key insight is to maintain an explicit mapping between MAT vertices and surface regions throughout the simplification process. 
Initially, the Voronoi diagram constructs the 3D medial mesh, which naturally induces a partition on the surface manifold (known as the Restricted Voronoi Diagram). This establishes a rigorous 2D-3D correspondence: each 3D medial vertex is intrinsically linked to several specific surface patches.
When a MAT edge is collapsed, the associated surface regions of the two endpoints are inherited by the new vertex. 
The optimal position of this new vertex is then determined by these associated surface regions, ensuring that the MAT remains anchored to the shape's bilateral symmetry. Simultaneously, we integrate a topological laplacian smoothing metric into the vertex positioning logic to promote a well-conditioned MAT mesh.

By leveraging these strategies, our surface-guided simplification framework maintains explicit correspondence between medial vertices and their associated surface regions throughout the entire process, enabling all decisions to directly reference the original geometry. This persistent correspondence allows our approach to produce clean, structure-aware medial axes with regularized, zigzag-free boundaries for both organic shapes and complex CAD models across varying levels of simplification; see Figure~\ref{fig:teaser}.

\begin{itemize}
    \item We develop a structural MAT framework that tracks the correspondence between medial vertices and surface regions, ensuring the MAT captures intrinsic surface symmetries and maintains structural alignment. 
    \item We propose a positioning strategy that jointly optimizes for geometric accuracy and triangulation quality, resulting in robust MAT meshes with clean boundaries across both organic and man-made geometries, with resilience to moderate levels of noise.
    \item We demonstrate the utility of our structural MAT in identifying the locus of rolling ball centers within fillet regions. Extracting this locus directly from discrete mesh representations is a significant challenge that, to our best knowledge, has not been addressed in the existing literature.
\end{itemize}

\section{Related Work}

The Medial Axis Transform (MAT) is a fundamental shape descriptor with extensive applications in shape approximation~\cite{petrov2024gem3d,yang2020p2mat}, simplification~\cite{yan2016erosion}, and analysis~\cite{lin2021point2skeleton}. Over the past decades, numerous methods have been proposed to approximate the 3D medial axis, each navigating the trade-offs between geometric accuracy, topological correctness, and computational efficiency. For comprehensive surveys on medial axis computation, we refer readers to~\cite{siddiqi2008medial,tagliasacchi20163d}. In this section, we review representative approaches most relevant to our work, categorized by their underlying methodology.

\subsection{Voronoi-based Medial Axis Computation}

The Voronoi diagram~\cite{https://doi.org/10.1111/j.1467-8659.2009.01521.x,10.1007/978-3-642-33573-0_21,10845125,levy2026graphitethree,11165079} is a primary tool for medial axis computation due to its spatial partitioning properties and inherent link to the medial axis. In 2D, if the boundary curve satisfies $\epsilon$-sampling conditions, the subset of the Voronoi diagram contained within the shape converges to the medial axis~\cite{brandt1994convergence}. However, this property does not directly extend to 3D due to the presence of ``slivers''---Voronoi vertices positioned infinitesimally close to the surface that produce redundant, unstable medial branches~\cite{amenta2001power}.

To address this, various filtering techniques have been developed. \textit{Angle-based filtering methods}~\cite{dey2004approximating,attali1996modeling} prune the Voronoi diagram using geometric criteria, though they frequently fail to guarantee topological preservation. The \textit{$\lambda$-medial axis}~\cite{chazal2005lambda,chazal2008smooth} discards medial spheres with radii smaller than a threshold $\lambda$; while this preserves homotopy, it tends to over-smooth fine geometric features. Alternatively, the \textit{Scale Axis Transform (SAT)}~\cite{giesen2009scale,miklos2010discrete} applies multiplicative scaling to medial spheres to remove unstable spikes while retaining small features by filtering balls that are ``devoured'' by their neighbors. Beyond filtering, \textit{Power Crust}~\cite{amenta2001power} utilizes the weighted Voronoi diagram of ``poles'' to establish connectivity, forming a power shape that converges to the medial axis as sampling density increases.

\subsection{Voxel-based Methods}

Rather than relying on discrete surface samples, voxel-based approaches discretize the target shape into a grid and extract medial subsets based on distance fields, such as Euclidean~\cite{hesselink2008euclidean,rumpf2002continuous} or chamfer distances~\cite{pudney1998distance}. \textit{Voxel Cores (VC)}~\cite{yan2018voxel} provides rigorous sampling conditions and computes the medial mesh as the dual of the interior Delaunay triangulation. While these methods offer theoretical guarantees of topological equivalence, they often face a trade-off between geometric precision and computational overhead: high accuracy requires extremely fine voxel resolutions, leading to significant memory consumption.

\subsection{PD-based and Patch-based Optimization}

Recent advances have moved toward using Power Diagrams (PD) for feature-aware MAT computation. \textit{MATFP}~\cite{wang2022matfp} employs surface Restricted Power Diagrams (RPD) to optimize Voronoi ball positions, ensuring they remain tangential to the boundary. It notably preserves external sharp edges by placing zero-radius spheres on non-smooth regions. \textit{MATTopo}~\cite{wang2024mattopo} further enhances topological robustness using volumetric RPDs, though the required tetrahedron slicing operations are computationally intensive. Both methods attempt to preserve internal features (seams and junctions) by inserting feature spheres when structural deficiencies are detected, which can occasionally lead to numerical instability or densely clustered primitives.

To improve mesh quality, \textit{MATStruct}~\cite{10.1145/3757377.3763840} introduces a particle-based energy optimization framework. However, the iterative recomputation of power diagrams within the optimization loop remains a major computational bottleneck. Similarly, patch-based methods~\cite{10845125} segment models into patches to compute generalized Voronoi diagrams. While effective for simple CAD models, these methods struggle when patch boundaries are ambiguous or when features between adjacent patches are not well-defined.

\subsection{Simplification and Sparse Approximation}

Given the density of raw medial extractions, several approaches focus on generating sparse representations. \textit{Q-MAT}~\cite{10.1145/2753755} employs iterative simplification via edge collapses guided by a quadric error metric. While flexible, Q-MAT suffers from significant ``reference drift''; because each step relies on the previous approximation rather than the original surface, geometric fidelity weakens over time, leading to irregular boundaries.

Conversely, \textit{skeletal point selection methods}~\cite{dou2022coverage,wang2024coverage} take a constructive approach by selecting a target number of skeletal points and inferring connectivity. \textit{VMAS}~\cite{10.1145/3680528.3687678} adopts a coarse-to-fine strategy by minimizing a hybrid metric. However, VMAS does not explicitly account for feature preservation---neither external sharp edges nor internal seams. Additionally, it can exhibit oscillatory behavior or instability when the sphere count exceeds a few hundred, preventing convergence to a clean, sparse representation. Our method addresses these limitations by leveraging the persistent correspondence between the 3D Voronoi diagram and the surface RVD, ensuring structural alignment even at extreme simplification levels.

\section{Problem Formulation and Overview}

\subsection{Problem Statement}
The Medial Axis Transform (MAT), as a compact and complete descriptor that encodes a shape’s topological structure and local thickness, plays a fundamental role in geometric representation.
Given a continuous or discrete boundary surface $\mathcal{S}$, the objective is to compute a medial structure $\mathbf{M}$, typically represented as a discrete mesh consisting of vertices, edges, and faces.
In engineering applications, this discrete outcome often serves as the basis for further transformation into quadrilateral representations or B-reps. Despite significant progress in this domain, several intrinsic challenges persist, which are further amplified when the input is a discrete mesh representation rather than a smooth manifold.

\paragraph{Requirements for a High-Quality MAT}
To characterize a high-quality MAT, we identify a set of often-conflicting requirements that a robust algorithm should satisfy:

\begin{enumerate}
    \item \textit{Geometric Fidelity}: The ability to accurately reconstruct the original surface $\mathcal{S}$ with minimal geometric deviation.
    \item \textit{Topological Inheritance}: The preservation of the topological properties, such as genus and homotopy type, of the original surface.
    \item \textit{Structural Alignment and Boundary Regularity}: The capacity to effectively capture intrinsic structural features---such as aligning boundaries with fillet centerlines---while maintaining clean, zigzag-free boundaries.
    \item \textit{Mesh Quality}: Ensuring well-conditioned triangle elements to avoid normal flips, near-zero area triangles, and other forms of mesh degeneracy.
    \item \textit{Conciseness and Scalability}: The ability to remain geometrically representative even at extremely low triangle counts, achieving ultra-low-poly representations.
    \item \textit{Computational Efficiency}: Optimized runtime performance suitable for processing large-scale or complex geometric inputs.
\end{enumerate}

Most prior research has prioritized reconstruction accuracy while paying less attention to structural alignment and boundary regularity. In this paper, our algorithm focuses on achieving high geometric fidelity and computational efficiency through a correspondence-preserving simplification framework. As a natural consequence of maintaining explicit medial-to-surface correspondence throughout simplification, our method also improves structural alignment and boundary regularity, while balancing the competing demands of mesh quality and conciseness. 
Regarding topological inheritance, we incorporate a heuristic protection strategy that helps reduce the risk of unintended topological changes during simplification, though strict preservation is not guaranteed for models of arbitrarily high genus.
\begin{figure}[h]
    \centering
    \includegraphics[width=.99\linewidth]{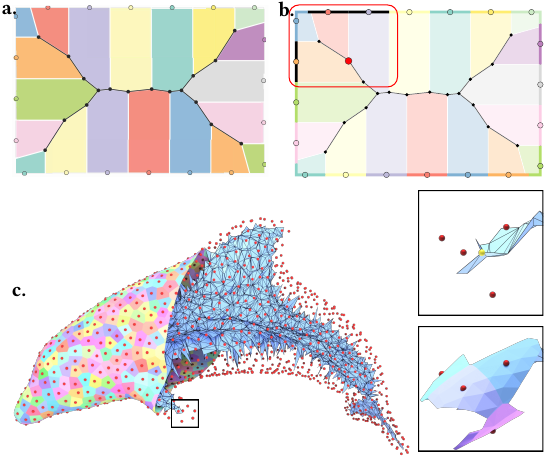}
    \caption{The geometric correspondence between the boundary surface and medial axis vertices. (a) 2D Construction: Surface samples generate the Voronoi diagram (MAT) and simultaneously induce the RVD on the boundary. (b) Local Correspondence: As highlighted, a single medial vertex is defined by three equidistant surface samples. We define the union of the three RVD cells belonging to these generating samples as the Atlas of the vertex, denoted as $\text{Atlas}(v_j)$. (c) 3D Duality: In our 3D framework, this concept generalizes: each medial vertex is defined by four surface samples, and its $\text{Atlas}(v_j)$ comprises the four corresponding surface cells.}
    \label{fig:voronoi_23d}
\end{figure}

\begin{figure*}[h]
    \centering
    \includegraphics[width=.99\linewidth]{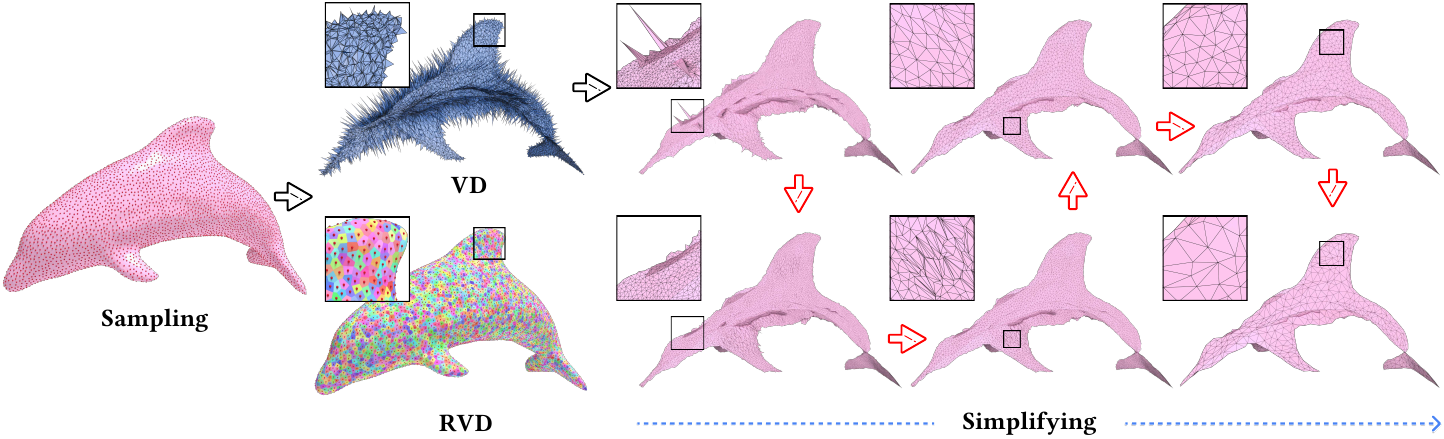}
    \caption{Pipeline overview. We begin by sampling the input surface and computing the 3D Voronoi diagram along with the Restricted Voronoi Diagram (RVD) on the surface.
    Each vertex in the 3D Voronoi diagram is assigned a set of surface regions, establishing a correspondence between the medial vertex and the mesh surface. This correspondence is inherited and updated throughout the simplification process.}
    \label{fig:pipeline}
\end{figure*}

\subsection{Overview}
A common pipeline for MAT computation begins with a set of surface samples $\{s_i\}_{i=1}^n$ extracted from~$\mathcal{S}$, where $\mathcal{S}$ is either a smooth surface or a discrete representation. This is followed by a pruning and simplification process of the Voronoi diagram $\mathcal{V}$ constructed from $\{s_i\}_{i=1}^n$.
Since the medial axis of a bounded shape lies strictly inside its boundary, only the inner portion of $\mathcal{V}$—comprising Voronoi vertices, edges, and faces enclosed by $\mathcal{S}$—is retained as the initial medial representation; all exterior components are discarded.
While previous methods often treat the simplification purely in the volumetric domain, we explicitly leverage the intrinsic geometric link between the dual structures: each 3D Voronoi cell naturally induces a corresponding region on the surface manifold (the Restricted Voronoi Diagram cell).
In our framework, we strictly enforce this correspondence, treating each surface sample $s_i$ as a proxy for this specific local surface region (see Figure~\ref{fig:voronoi_23d}(a)).

In a general configuration, each Voronoi vertex~$v_j \in \mathcal{V}$ is equidistant to three boundary samples in 2D or four surface samples in 3D. Consequently, $v_j$ can be regarded as being jointly determined by three boundary segments in 2D or four surface regions in 3D, as illustrated in Figure~\ref{fig:voronoi_23d}(b,c). We define the set
\begin{equation}
\text{Atlas}(v_j) = \{ \text{Cell}(s_i) \mid s_i \text{ is a contributing site for } v_j \}
\end{equation}
to denote the collection of surface regions associated with a Voronoi vertex~$v_j$. This set represents the fundamental geometric correspondence between a medial-axis primitive (i.e., a mesh vertex of the current medial-axis surface) and the boundary manifold.

This intrinsic duality inspires our framework to progressively simplify the MAT while explicitly maintaining this correspondence throughout the entire pipeline. Specifically, when an edge~$e$ is collapsed to form a new vertex~$v$, we infer the optimal position of $v$ based on the aggregated surface regions inherited from its two endpoints. Furthermore, we incorporate constraints for triangle quality and feature line alignment when determining the position of $v$. An overview of our pipeline is illustrated in Figure~\ref{fig:pipeline}.

\section{Method}
Our method takes a triangulated surface mesh as input and produces a simplified, high-quality medial mesh through surface-guided optimization. The algorithm begins by sampling the input surface and computing dual Voronoi structures: the inner portion of a 3D Voronoi diagram in volumetric space, which serves as the initial medial representation, and a Restricted Voronoi Diagram (RVD) on the surface manifold, which provides geometric guidance throughout simplification. We then progressively refine this initial structure through iterative edge collapses, with each operation strictly anchored to the original input surface via the established correspondence.

The remainder of this section is organized as follows. 
Section~\ref{sec:Initialization} describes the construction of the dual Voronoi structures and introduces a medial face filtering mechanism designed to eliminate spurious faces near concave features. 
Section~\ref{sec:OptimalVertexPlacement} details the optimization framework for edge collapse, where we determine the optimal position of new vertices by minimizing a composite energy functional that balances surface fidelity with mesh quality. 
Section~\ref{sec:EdgePrioritization} presents our edge prioritization strategy, which ranks collapse operations based on a unified metric combining geometric error, element quality, and structural stability. 
Finally, Section~\ref{sec:CADFeaturePreservation} outlines the specialized handling for CAD models, ensuring that sharp features and corners are accurately preserved throughout the simplification process.

\subsection{Initialization}
\label{sec:Initialization}
\paragraph{Surface Sampling and Voronoi Construction} 
We begin by performing blue noise sampling on the input surface $\mathcal{S}$ to obtain a dense set of surface samples, typically ranging from $10\text{K}$ to $100\text{K}$ points. These samples serve as the generating sites for a 3D Voronoi diagram $\mathcal{V}$. 
The Restricted Voronoi Diagram (RVD) is defined as the restriction of $\mathcal{V}$ to the surface $\mathcal{S}$, where each RVD cell is given by:
\begin{equation}
    \text{Cell}(s) = \mathcal{V}(s) \cap \mathcal{S},
\end{equation}
where $\mathcal{V}(s)$ is the Voronoi cell of $s$ in $\mathbb{R}^3$. This yields a partition of $\mathcal{S}$ into surface regions, each associated with a unique generating sample.

In a general configuration, each Voronoi vertex $v$ in $\mathcal{V}$ is equidistant to exactly four surface samples, denoted as the set $S_v$. 
This duality between Voronoi vertices and surface regions is the key structure we exploit. We define the \textit{Atlas} of vertex $v$ as the union of the RVD cells belonging to its four generating sites:
\begin{equation}
\text{Atlas}(v) = \bigcup_{s \in S_v} \text{Cell}(s).
\end{equation}
This establishes an explicit correspondence between each medial vertex $v$ and a specific region of the input surface, effectively encoding the local bilateral symmetry of the shape (see Figure~\ref{fig:voronoi_23d}(c)).

\paragraph{Medial Face Filtering near Concave Features} 
In our dual Voronoi framework, each face in the 3D Voronoi diagram corresponds to two surface sample sites whose Voronoi cells share that interface. When both sites lie near concave feature lines, the medial face becomes geometrically ambiguous—it may represent a valid medial structure or merely a spurious artifact induced by the concave feature. Standard energy-based edge collapse cannot reliably distinguish and eliminate these questionable structures.

\begin{figure}[htbp]
    \centering
    \includegraphics[width=.99\linewidth]{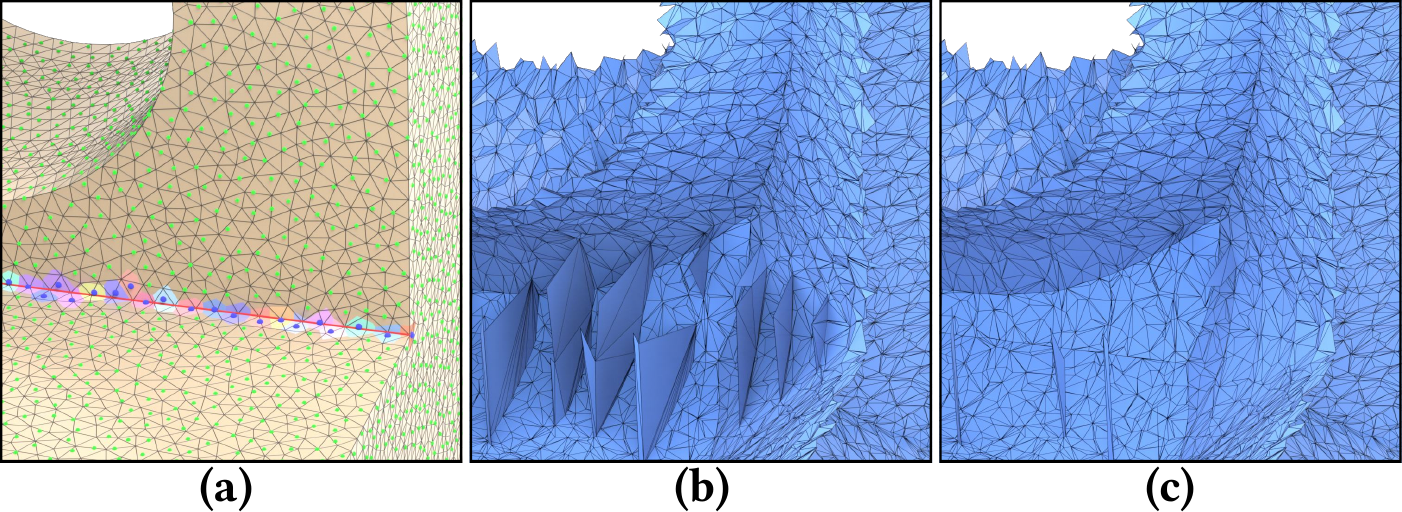}
    \caption{Medial face filtering near concave features. (a) Concave features on the CAD model are marked with red lines. Cells intersecting these feature lines are highlighted, with their corresponding Voronoi sites marked in blue. (b) Initial medial structures often include spurious faces where both endpoints lie on the same concave feature line. (c) The medial structure after filtering removes faces that fail the midpoint test, resulting in a cleaner medial topology in concave regions.}
    \label{fig:ConcaveFaceDelete}
\end{figure}
We therefore apply a geometric test to filter such faces before simplification begins. 
This filtering is strictly restricted to medial faces generated by two sites whose respective Restricted Voronoi cells both intersect concave feature lines.
For such a medial face $f$ corresponding to surface sites $s_1$ and $s_2$, we evaluate the midpoint $m = (s_1 + s_2)/2$. 
Our heuristic follows from a simple geometric observation: if the face generated by $s_1$ and $s_2$ truly lies on the medial axis, then its midpoint $m$ must also lie on the medial axis.
Consequently, $m$ should satisfy the empty ball property, meaning its distance to the surface is determined solely by $s_1$ and $s_2$.
We identify and remove spurious faces that violate this condition by checking:
\begin{equation}
    d_{\mathcal{S}}(m) < \alpha \cdot \frac{\|s_1 - s_2\|}{2},
    \label{eq:concave:filter}
\end{equation}
where $d_{\mathcal{S}}(m)$ is the distance from $m$ to the closest surface point, and $\alpha = 0.7$. 
If this inequality holds, it implies that $m$ fails to lie on the medial axis, and therefore, the face $f$ itself cannot be part of the valid medial axis.
This preprocessing removes spurious structures near concave features, enabling more reliable subsequent simplification. Figure~\ref{fig:ConcaveFaceDelete} shows the medial structure in concave regions before and after this filtering step, demonstrating the removal of spurious faces.

\subsection{Optimal Vertex Placement}
\label{sec:OptimalVertexPlacement}

The operation of collapsing a medial edge $e \triangleq v_1 v_2$ into a single vertex $v$ involves two primary considerations: the geometric fidelity of the reconstructed boundary surface and the mesh quality of the medial surface triangulation. To determine the optimal target position $v_{\text{new}}$ for an edge collapse, we introduce a composite energy functional consisting of two terms:
\begin{equation}
E(e \rightarrow v) = E_{\text{Fidelity}}(e \rightarrow v) + \lambda E_{\text{Lap}}(e \rightarrow v),
\label{eq:total:energy}
\end{equation}
where $v$ represents the candidate position for the new vertex, $E_{\text{Fidelity}}$ penalizes geometric deviation from the original boundary, and $E_{\text{Lap}}$ promotes well-conditioned triangle elements. The parameter $\lambda$ is a weighting coefficient that balances these requirements. The optimal position is defined as the minimizer of this functional:
\begin{equation}
v_{\text{new}} = \arg\min_{v} E(e \rightarrow v).
\end{equation}
We denote the aggregated surface regions as $\mathcal{A} = \text{Atlas}(v_1) \cup \text{Atlas}(v_2)$, inherited from the edge endpoints. 
Upon collapse, the target vertex $v_{new}$ inherits this correspondence, meaning $\mathcal{A}$ becomes the associated surface region for $v_{new}$ and is continuously propagated through subsequent simplification steps. This ensures that every medial vertex remains linked to a specific patch of the original input surface.

Each Restricted Voronoi Diagram (RVD) cell within $\mathcal{A}$ is classified as either \textit{invaginated} or \textit{regular}, depending on whether it intersects concave features.
This classification is necessary because these two types of cells exhibit fundamentally different behaviors during optimization, requiring distinct energy formulations.

\paragraph{Invaginated vs. Regular Cells} 
For a point on a smooth surface region, the local tangent plane defined by the point and its normal generally provides a good approximation to the original surface in the point's neighborhood. Existing methods exploit this approximation by using the distance from a spatial location to the tangent plane as a proxy for the true distance to the surface. One of the most widely used techniques based on this principle is the Spherical Quadric Error Metric (SQEM)~\cite{thiery2013sphere}.

\setlength{\columnsep}{5pt} 
\begin{wrapfigure}{r}{0.4\linewidth} 
\begin{center} 
\includegraphics[width=0.99\linewidth]{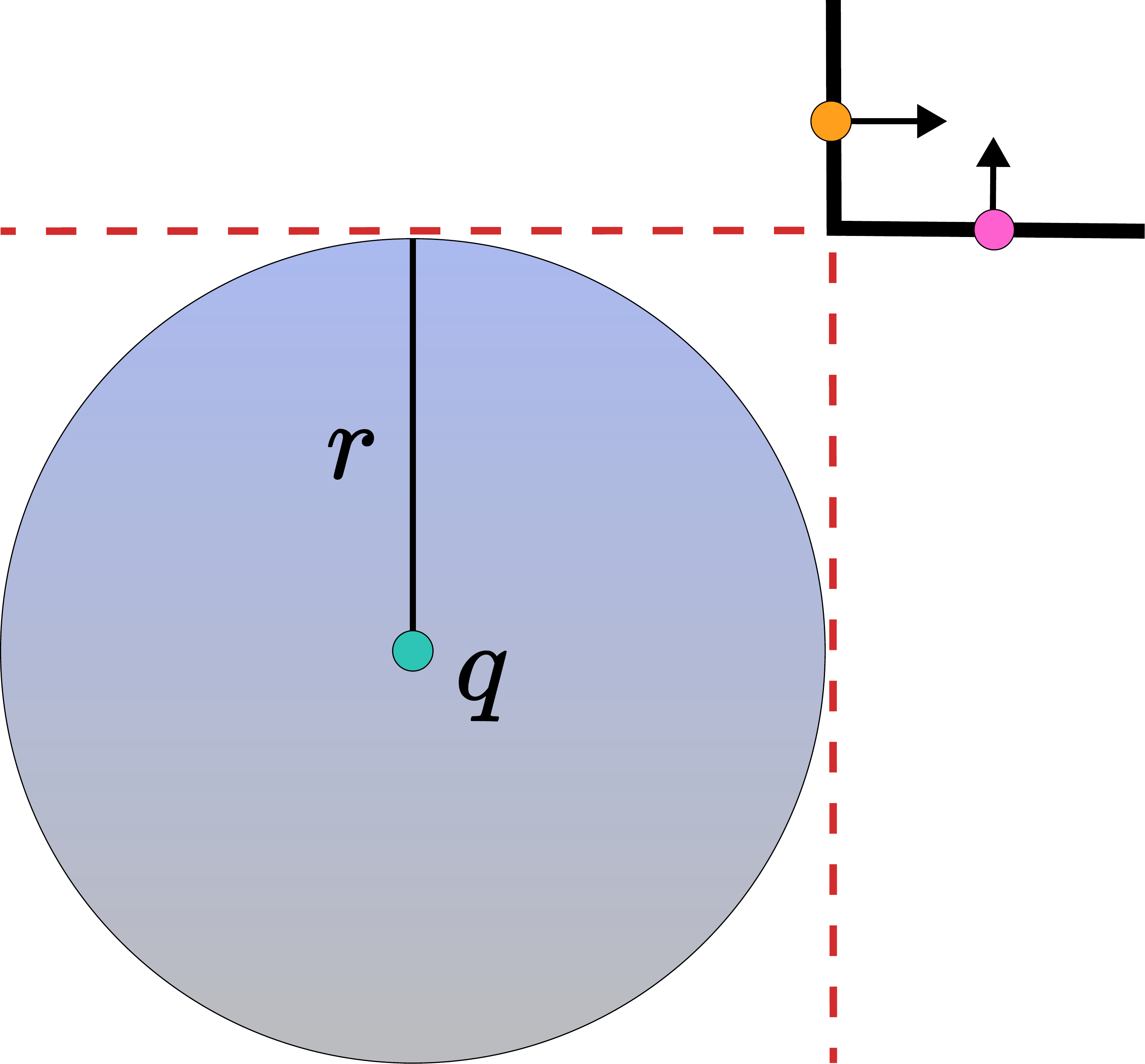} 
\end{center} 
\end{wrapfigure}
However, this tangent plane approximation breaks down near concave features. When sample points lie close to concave (invaginated) feature lines, the distance to the tangent plane can deviate significantly from the true distance to the surface. As illustrated in the inset, the signed distance from the medial sphere to the tangent plane fails to approximate the true distance to the boundary, rendering SQEM-based optimization unreliable in these regions.

To handle this issue, we classify each RVD cell based on its proximity to concave features. We detect concave features using dihedral angle thresholds: a surface edge is classified as concave if its dihedral angle exceeds $\pi + \phi$, where $\phi$ is a user-specified tolerance parameter. A cell $\text{Cell}_i$ is marked as an \textit{invaginated cell} if it intersects any concave feature line; otherwise, it is classified as a regular cell. 
Figure~\ref{fig:ConcaveFaceDelete}(a) illustrates the identified \textit{invaginated cells} and their corresponding surface sites on a local region of a CAD model.
This classification determines the appropriate optimization strategy for vertex placement during edge collapses.

\paragraph{Fidelity Term for Regular Cells} 
\begin{wrapfigure}{r}{0.4\linewidth} 
\begin{center} \includegraphics[width=0.99\linewidth]{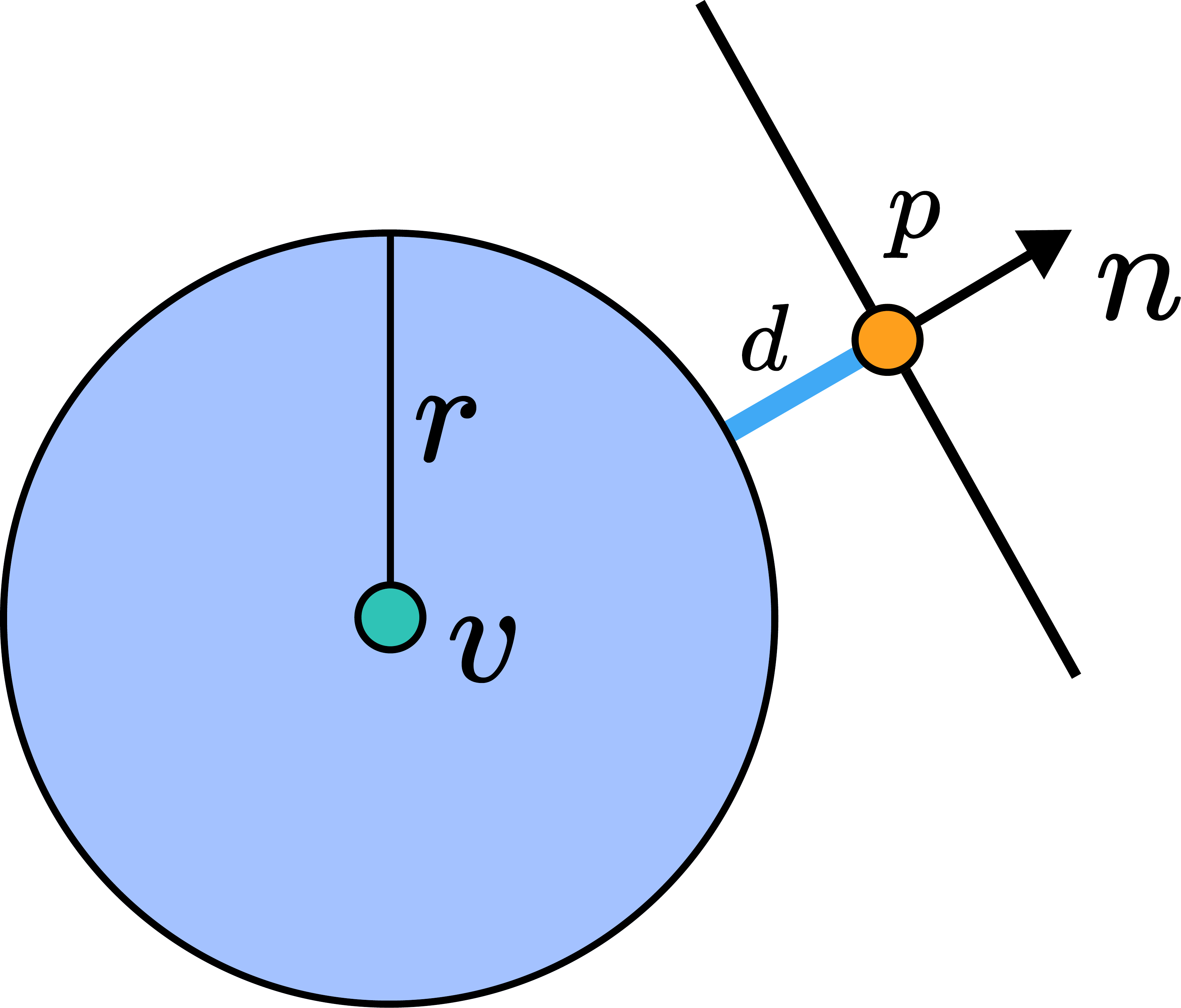} 
\end{center} 
\end{wrapfigure}
For a regular cell (one without concave features), we formulate its fidelity energy using the Spherical Quadric Error Metric (SQEM) framework~\cite{thiery2013sphere}, which measures the distance between a medial sphere and surface elements. For a medial sphere $s = (v, r)$ with center $v$ and radius $r$, the signed distance to a surface point $p$ with normal $n$ is: \begin{equation}
    d_{p,\mathbf{n}}(v,r) = \mathbf{n}^\top (p - v) - r.
\end{equation}

By defining the augmented state vector $\overline{v} = (v_x, v_y, v_z, r)^\top \in \mathbb{R}^4$, the squared distance can be expressed as a quadratic form: 
\begin{equation}
    d^2_{p,\mathbf{n}}(v,r) = \overline{v}^\top \mathbf{A}_p \overline{v} + \mathbf{b}_p^\top \overline{v} + c_p.
\end{equation}
where $A_p$, $b_p$, and $c_p$ are coefficients determined by point $p$ and normal $n$. 
The fidelity energy for a single cell $\text{Cell}_i$ is the integral of this metric over the cell region: 
\begin{equation} 
E_{\text{Fidelity}}(e \rightarrow v) \big|_{\text{Cell}i} = \int_{p \in \text{Cell}_i} \left(\overline{v}^\top \mathbf{A}_p \overline{v} + \mathbf{b}_p^\top \overline{v} + c_p \right)dA. 
\end{equation}

Since the surface region within each cell is composed of multiple planar polygons, the integral over the entire cell can be efficiently computed as the sum of area-weighted quadric terms from these individual facets.
The quadratic coefficients for each cell can be pre-computed during initialization and accumulated through matrix additions during simplification.

\paragraph{Fidelity term for invaginated RVD cells} 
\setlength{\columnsep}{5pt}

When an RVD cell $\text{Cell}_i$ is invaginated, SQEM becomes problematic due to significant variations in surface normals within the patch. In such cases, we move away from normal-based distance and instead treat $\text{Cell}_i$ as a geometric anchor for the medial sphere. We modify the formulation as follows:
\begin{equation}
    E_{\text{Fidelity}}(e \rightarrow v) \big|_{\text{Cell}_i} = \int_{p \in \text{Cell}_i} \left( \|v - p\| - r \right)^2 dA.
    \label{eq:Fidelity:non-convex}
\end{equation}
To optimize for computational efficiency, we further relax this objective by using the cell's generating site $s_i$ as a representative proxy for the patch:
\begin{equation}
    E_{\text{Fidelity}}(e \rightarrow v) \big|_{\text{Cell}_i} \approx \text{Area}(\text{Cell}_i) \cdot \left( \|v - s_i\| - r \right)^2,
    \label{eq:Fidelity:non-convex:2}
\end{equation}
where $s_i$ is the site corresponding to the invaginated RVD cell $\text{Cell}_i$. 
This formulation effectively stabilizes vertex placement within invaginated regions.

\paragraph{Topological Laplacian smoothing term}
When a medial edge $e \triangleq v_1 v_2$ is collapsed into a vertex $v$, let $\mathcal{N}(v_1, v_2)$ denote the union of the neighboring vertices of $v_1$ and $v_2$ (excluding $v_1$ and $v_2$ themselves). 
We incorporate topological (uniform) Laplacian smoothing term over the merged neighborhood to promote well-conditioned triangulation: 
\begin{equation}
    E_{\text{Lap}}(e \rightarrow v) = \frac{1}{|\mathcal{N}(v_1, v_2)|} \sum_{u \in \mathcal{N}(v_1, v_2)} \| v - u \|^2,
    \label{equ:TQuality}
\end{equation}
where $|\mathcal{N}(v_1, v_2)|$ denotes the cardinality of the neighbor set.
Minimizing this term encourages the target vertex to be positioned near the centroid of its neighborhood, promoting local isotropy and the formation of near-equilateral triangles, which significantly improves the triangle quality and numerical stability of the simplified medial mesh for downstream applications.

\paragraph{Combined Optimization} 
The total energy combines fidelity contributions from all cells with a laplacian smoothing term:
\begin{equation}
E(e \rightarrow v) = \sum_{\text{Cell}_i \in \mathcal{A}} E_{\text{Fidelity}}(e \rightarrow v) \big|_{\text{Cell}_i}+\lambda E_{\text{Lap}}(e \rightarrow v)
\end{equation}

The optimization strategy depends on whether $\mathcal{A}$ contains \textit{invaginated cells}. For regions without \textit{invaginated cells}, both $E_{\text{Fidelity}}$ (SQEM) and $E_{\text{Lap}}$ are quadratic, reducing the optimization to solving a small linear system with a closed-form solution. 
For regions with invaginated cells, the square root term in Equation~\ref{eq:Fidelity:non-convex:2} makes the energy nonlinear, which we solve using L-BFGS optimization~\cite{Liu1989OnTL}. In both the linear and nonlinear cases, the optimization variables are the four degrees of freedom of the candidate medial sphere, namely the center coordinates $v = (v_x, v_y, v_z)$ and the radius $r$. All other quantities—including the surface sites $s_i$, the surface points $p$ with their normals $\mathbf{n}$, and the associated cell regions in $\mathcal{A}$—are treated as fixed constants inherited from the input surface at the moment of edge collapse, and are not updated during the optimization. In particular, $\mathcal{A}$ is determined once by the union of the Atlases of the two endpoints and remains unchanged while the L-BFGS iterations refine $(v, r)$.

To ensure stability in the nonlinear case, we initialize $v$ at the edge midpoint $\frac{v_1 + v_2}{2}$ and set $r$ as the minimum distance from this midpoint to all surface sites in $\mathcal{A}$. This geometrically meaningful initialization improves convergence. The weight $\lambda$ balances surface fidelity and mesh quality across both formulations.

The explicit gradient formulas with respect to $(v, r)$ for both regular and invaginated cells are derived in Appendix~\ref{app:gradient}.

\subsection{Priority for Edge Collapse}
\label{sec:EdgePrioritization}

The energy functional defined in Section~\ref{sec:OptimalVertexPlacement} serves a dual purpose. While its minimization yields the optimal target configuration, the resulting minimum energy value $E(e \rightarrow v_{\text{new}})$ provides a quantitative measure of the geometric deviation introduced by the collapse. In a standard simplification framework, one would typically prioritize collapsing edges with the lowest reconstruction cost $E$ to maximize fidelity.

However, for medial axis simplification, relying solely on geometric error is insufficient. A critical requirement is the effective removal of unstable branches ("spikes")—artifacts sensitive to boundary noise—before simplifying the stable structural components. To address this, we incorporate a structural stability measure into the prioritization logic.

To quantify structural stability, we directly adopt the stability measure from Q-MAT~\cite{10.1145/2753755}. For an edge $e=(v_1, v_2)$ with radii $r_1$ and $r_2$, the stability ratio is defined as:
\begin{equation}
\text{Spike}(e) = \frac{\max\{0, \|v_1 - v_2\| - |r_1 - r_2|\}}{\|v_1 - v_2\|}.
\end{equation}
This normalized metric ranges from $0$ (representing a pure spike) to $1$ (representing a stable edge).

To unify these two considerations into a single scalar for prioritization, we define a cost function:
\begin{equation}
    \text{Cost}(e) = E(e \rightarrow v_{\text{new}}) \cdot \Psi(\text{Spike}(e)),
    \label{equ:Priority}
\end{equation}
where $\Psi(x)$ is a sigmoid-based weighting function:
\begin{equation}
    \Psi(x) = \frac{1}{1 + e^{-k(x - \tau)}}.
\end{equation}

\begin{wrapfigure}{r}{0.35\linewidth}
 \centering
 \includegraphics[width=0.99\linewidth]{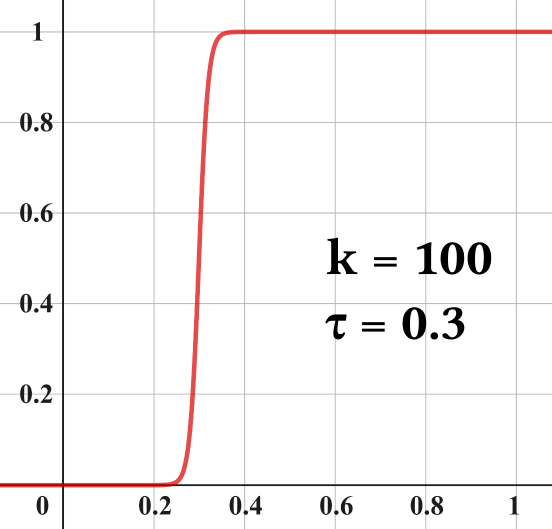}
\end{wrapfigure}
In this formulation, $\tau$ acts as the stability threshold. We employ a large sharpness parameter (e.g., $k=100$) to enforce a near-binary transition behavior:
\begin{itemize}\item \textbf{Spike Removal:} For unstable edges ($\text{Spike}(e) < \tau$), $\Psi$ vanishes, forcing their immediate removal regardless of geometric error.\item \textbf{Geometry Simplification:} For stable branches ($\text{Spike}(e) > \tau$), $\Psi$ saturates to 1, ensuring the collapse order is dominated purely by geometric fidelity.\end{itemize}
All candidate edges are maintained in a priority queue ordered by this weighted cost.
The process iteratively collapses the least-cost edge and updates the neighborhood until the target complexity is reached.

Furthermore, to help preserve topological fidelity during the coarse simplification phase, we adopt a protection strategy similar to~\cite{10.1145/2753755}. Specifically, once the vertex count drops below 200, we enforce the \textit{Link Condition}~\cite{Dey1998TopologyPE} prior to every edge collapse. This constraint explicitly identifies and forbids contractions that would alter the mesh topology—such as closing essential holes or collapsing tunnels—helping to reduce the risk of topological changes during simplification. It should be noted that this strategy is heuristic in nature: on models with highly complex topology, topological errors may still occur, particularly when the target vertex count is insufficient to faithfully represent the input genus.

\begin{figure}[h]
	\centering
\includegraphics
[width=.99\linewidth]{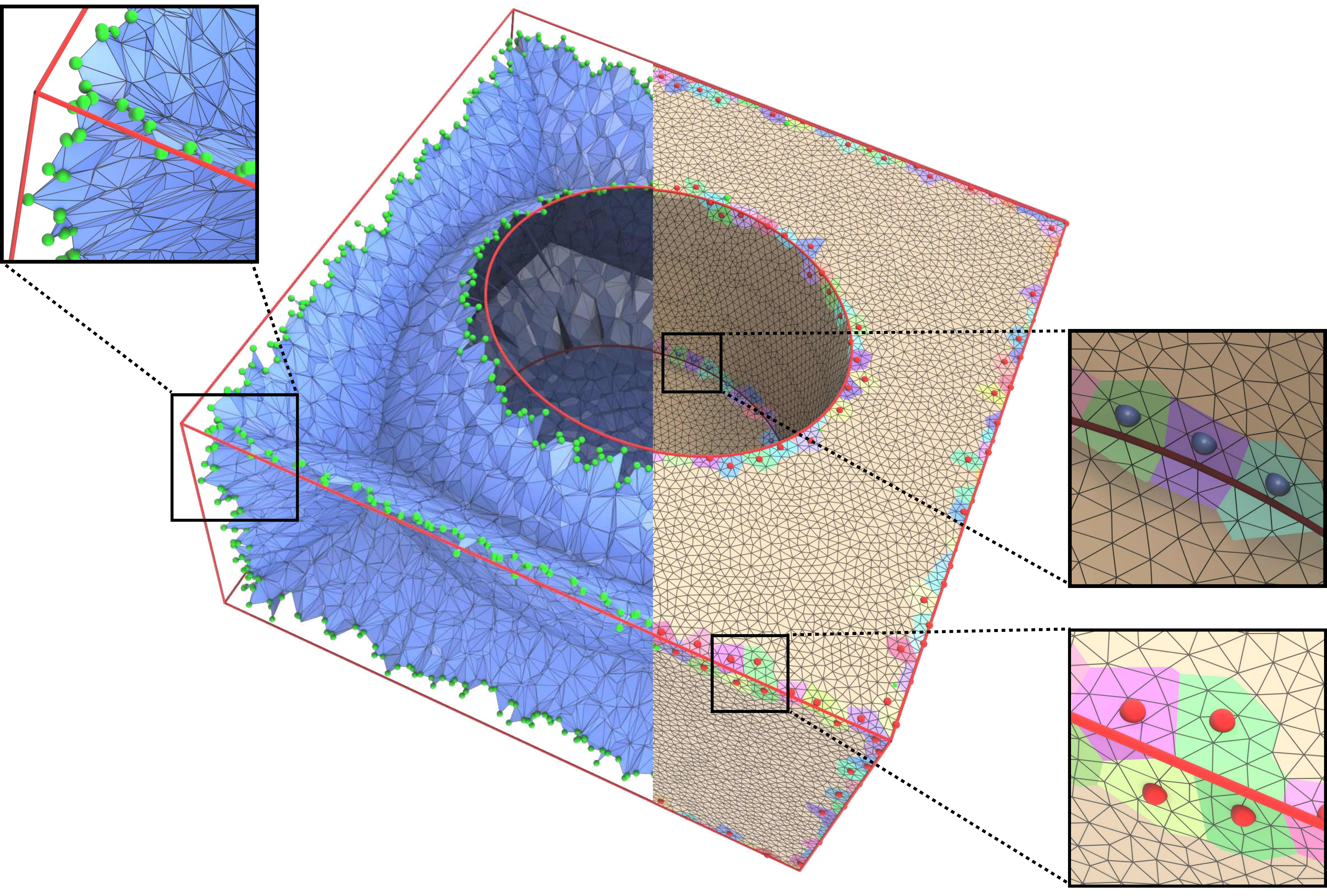}
\caption{Feature classification. Convex features and concave features on the model surface are marked in red and maroon, respectively. All Restricted Voronoi cells that intersect feature lines are highlighted, along with their corresponding sample sites (see right close-ups). Additionally, feature vertices are marked in green (left close-up).}
\label{fig:ConvexConcaveFeature}
\end{figure}

\subsection{Extension to CAD Models}
\label{sec:CADFeaturePreservation}

CAD models contain numerous sharp features such as edges and corners. 
A fundamental characteristic of the MAT for such shapes is that the medial sheets extend outward and terminate precisely at convex feature lines, where the medial radius vanishes.
To accurately capture and preserve these geometric details (which are often lost in smooth approximations), we introduce additional preprocessing and optimization strategies.

\paragraph{Feature Classification}
Specifically, we define a surface edge as a \textit{sharp feature} if its dihedral angle is less than $\pi - \phi$, where $\phi$ is a user-specified tolerance. Accordingly, any RVD cell intersecting such a sharp feature line is designated as a \textit{feature cell}.

This cell-level characterization enables us to identify \textit{feature vertices} in the medial mesh. A medial vertex $v$ is classified as a \textit{feature vertex} if its associated surface region $Atlas(v)$ contains at least one \textit{feature cell} but is entirely free of \textit{invaginated cells}. This criterion isolates stable, prominent corners or ridges for preservation, while explicitly excluding unstable geometry associated with invaginated regions. Figure~\ref{fig:ConvexConcaveFeature} illustrates these categories and their geometric correspondence.

\begin{figure}[h]
    \centering
    \includegraphics[width=.99\linewidth]{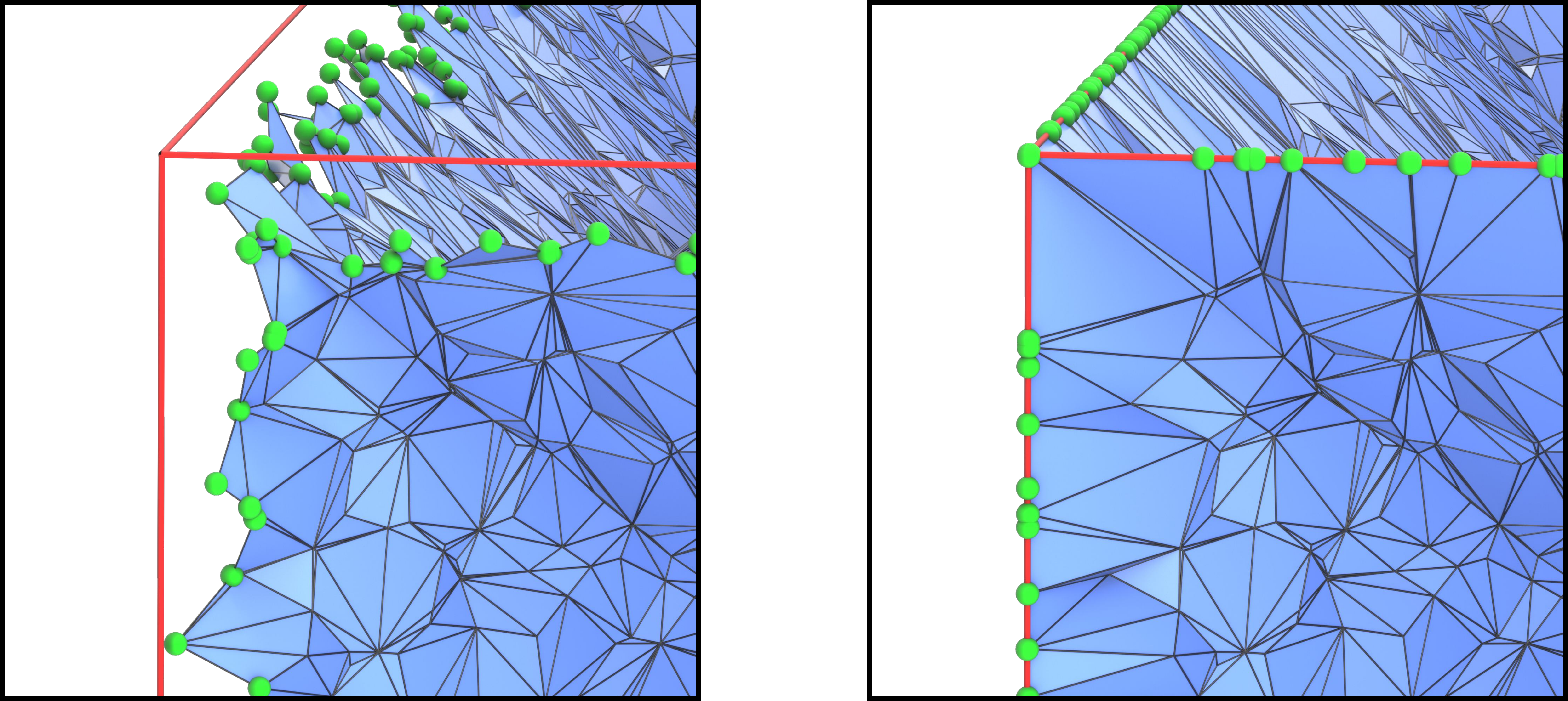}
    \caption{Comparison of \textit{feature vertices} before (left) and after (right) being snapped to convex feature lines. By enforcing $r=0$, we prevent the medial boundary from drifting and ensure accurate capture of sharp CAD features.}
    \label{fig:ConvexSnap}
\end{figure}

\paragraph{Feature Snapping.}
Prior to simplification, we perform a preprocessing step to align \textit{feature vertices} with convex boundary features.
We adapt the energy functional defined in Equation~\ref{eq:total:energy} to a single-vertex setting.
Crucially, we enforce a zero-radius constraint ($r=0$) to drive the medial vertex onto the boundary. The optimization problem is formulated as:
\begin{equation}
v_{\text{snap}} = \arg\min_{v} \left( E_{\text{Fidelity}}(v, r=0) \big|_{\text{Atlas}(v)} + \lambda E_{\text{Lap}}(v) \big|_{\mathcal{N}(v)} \right).
\end{equation}
By fixing $r=0$, the fidelity term reduces to minimizing the squared distance from $v$ to the surface patch $\text{Atlas}(v)$, which is equivalent to the classical QEM formulation of~\cite{10.1145/258734.258849} in this setting. This naturally pulls $v$ toward the sharp feature line where the medial radius vanishes.
This effectively ``snaps'' the medial boundary to the CAD model's sharp features, as illustrated in Figure~\ref{fig:ConvexSnap}, ensuring accurate capture of geometric details before simplification begins.

\paragraph{Feature Edge Collapse}
To ensure that feature vertices remain strictly anchored to convex boundary features during simplification, we adopt a discrete candidate sampling strategy for edges incident to feature vertices (provided no invaginated RVD cells are involved).
This choice is motivated by the need to prevent excessive drifting along feature lines and ensure numerical stability when the zero-radius constraint is enforced. Specifically, we select the target $v$ from a finite set of candidates $\mathcal{X}$.
If only one endpoint is a \textit{feature vertex}, we restrict the candidate set to that vertex alone ($\mathcal{X} = \{v_{\text{feat}}\}$). If both are feature vertices, we consider both endpoints and their midpoint: $\mathcal{X} = \{v_1, v_2, (v_1+v_2)/2\}$.

For each candidate $x \in \mathcal{X}$, we enforce a zero-radius constraint ($r=0$) to preserve sharp features. The collapse cost is evaluated by adapting the general energy functional (Equation~\ref{equ:Priority}):
\begin{equation}
C_{\text{feature}}(e, x) = \left( E_{\text{Fidelity}}(x, 0) \big|_{\mathcal{A}} + \lambda E_{\text{Lap}}(x) \big|_{\mathcal{N}} \right) \cdot \Psi(\text{Spike}(e)),
\end{equation}
where $\mathcal{A} = \text{Atlas}(v_1) \cup \text{Atlas}(v_2)$ is the aggregated surface region. The term $\Psi(\text{Spike}(e))$ is a weighting factor derived from the edge stability classification (as defined in Section~\ref{sec:EdgePrioritization}), used to prioritize stable collapses. The candidate $x^*$ minimizing this cost is selected as the collapse target, and the cost determines the edge's priority in the simplification queue.

\begin{figure*}[h]
	\centering
\includegraphics
[width=0.99\linewidth]{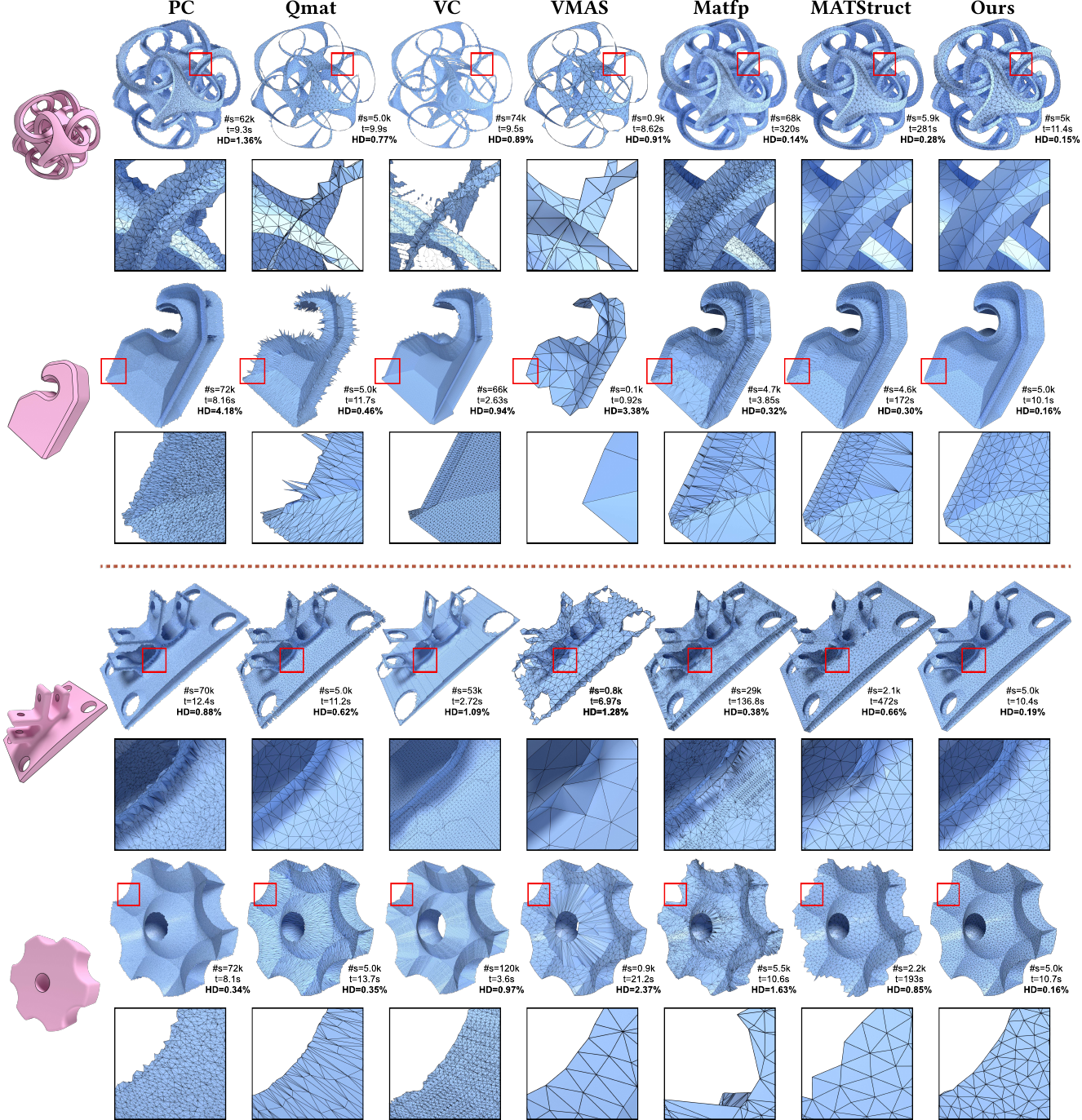}
\caption{Medial axis comparisons on representative CAD models with sharp features (top two rows) and regions with smooth transitions (bottom two rows). Results include computation time ($t$), sphere count ($\#s$), and the Hausdorff distance (HD) between the reconstructed surface and the original input. 
Our surface-guided approach accurately captures both sharp feature alignments and centerlines of smooth transitions.
}
\label{fig:CADComparasion}
\end{figure*}
\section{Evaluation}
\textbf{Experimental Setup.}
In this section, we present quantitative and qualitative evaluations of the proposed method. We implemented our algorithm in C++ and conducted experiments on two platforms: a Mac Mini with M4 chip and 16GB RAM served as our primary testing platform, while comparisons with methods requiring CUDA or Windows were performed on a system with Intel i7-14700K CPU, 64GB RAM, and NVIDIA RTX 4090 GPU.
Our implementation uses SurfaceVoronoi~\cite{10.1145/3550454.3555453} for RVD computation, CGAL~\cite{cgal:pt-t3-25b} for 3D Voronoi diagram construction, and FCPW~\cite{FCPW} for efficient point-to-mesh distance queries.

All input models were normalized to the unit cube $[0,1]^3$ during testing. We set the algorithm parameters as follows: $\lambda = 6 \times 10^{-6}$ and $\tau = 0.025$. For surface sampling, we employed blue noise sampling. In comparative evaluations, we used $10K$ sample points and simplified the medial axis to $1K$ vertices for freeform models, and $50K$ sample points with simplification to $5K$ vertices for CAD models. Our test models were sourced from Thingi10K~\cite{Thingi10K}, the ABC Dataset~\cite{Koch_2019_CVPR}, and~\cite{10.1145/3680528.3687678}.
We compared our method against seven state-of-the-art approaches: Q-MAT~\cite{10.1145/2753755}, VC (Voxel Cores)~\cite{yan2018voxel}, PC (Power Crust)~\cite{amenta2001power}, VMAS~\cite{10.1145/3680528.3687678}, MATFP~\cite{wang2022matfp}, MATStruct~\cite{10.1145/3757377.3763840}, and~\cite{10845125}. Among these, MATStruct and VC were evaluated on the Windows platform, while all other methods were tested on the Mac platform.

\textbf{Evaluation Metrics.} We evaluate the reconstruction quality of the medial axis using bidirectional Hausdorff distance (HD) between the input surface and the reconstructed surface. All Hausdorff distances are reported as percentages relative to the diagonal length of the model's bounding box. Surface reconstruction from the medial axis is performed using the Blender MAT addon~\cite{blender-mat-addon}. In the comparative tables, $\#s$ denotes the number of medial spheres in the simplified medial mesh, and $t$ represents the computation time in seconds.

\textbf{Special Note on VMAS.} 
Since the effective capacity of VMAS is inherently tied to the density of the initial sampling, its performance varies with the input configuration. In our experiments, the input meshes contain approximately $15\text{K}$ vertices on average.
VMAS runs rapidly but does not terminate automatically, and the optimization may become unstable as the target sphere count increases. For VMAS, the reported time $t$ reflects the duration until we manually stopped the optimization after observing stable convergence.
All timings for VMAS are reported with artificial slow-down disabled for a fair benchmark.

\begin{figure}[h]
	\centering
\includegraphics
[width=0.99\linewidth]{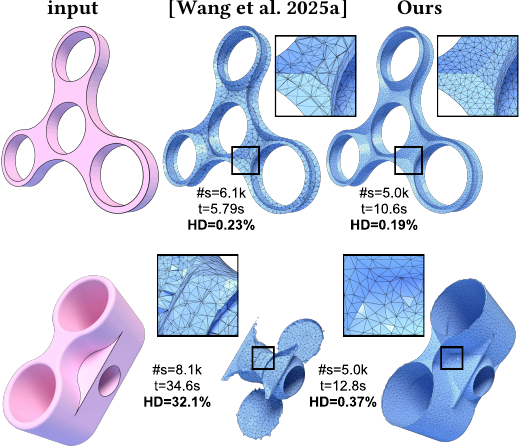}
\caption{Comparison with~\cite{10845125} on CAD models. \cite{10845125} produces reasonable results when models can be cleanly segmented into patches (top models), but struggles or fails entirely when faced with models that resist straightforward patch decomposition (bottom models). Our method handles both categories uniformly without requiring segmentation preprocessing.}
\label{fig:spaceVoronoi}
\end{figure}
\subsection{Comparisons on CAD Models}
We evaluate our method on two representative categories of CAD models: models with sharp, well-defined features that can be detected via dihedral angle thresholds, and models containing smoothly blended edges where features gradually transition without clear boundaries. While sharp features can be identified through preprocessing, smooth transitions present a more challenging test—the true medial axis should trace the centerline of the transitional region. An algorithm's ability to accurately capture these centerlines reflects its capacity to faithfully represent the overall surface geometry.

Figure~\ref{fig:CADComparasion} shows comparative results on both categories, including computation time ($t$), number of medial spheres ($\#s$), and bidirectional Hausdorff distance (HD). Methods such as PC, Q-MAT, VC, and VMAS were not designed with CAD-specific handling in mind, and consequently struggle to capture sharp features—their results exhibit irregular, jagged boundaries that fail to align with the geometric features of the input models.

MATFP~\cite{wang2022matfp} and MATStruct~\cite{10.1145/3757377.3763840} represent state-of-the-art approaches specifically designed for CAD models. For models with sharp features, both methods demonstrate clear advantages over general-purpose approaches: their medial boundaries are better aligned with surface features, marking a significant step forward. 
However, both methods also exhibit certain limitations. 
MATFP produces meshes with suboptimal triangle quality, while MATStruct achieves better triangulation at the cost of occasionally missing narrow feature regions (see the first model in Figure~\ref{fig:CADComparasion}). Additionally, MATStruct requires iterative power diagram recomputation, leading to slow performance and GPU dependency that limits its applicability. 
Both methods also exhibit limitations in capturing internal medial features—in some cases, the medial structure deviates from expected straight feature lines, appearing curved or irregular where geometric features should be linear (observe the second model in Figure~\ref{fig:CADComparasion}).
For models containing smooth feature transitions, the limited surface awareness of both methods hinders their ability to accurately trace the centerlines through these gradual changes, resulting in irregular medial boundaries that fail to capture the smooth transitional regions.

Figure~\ref{fig:spaceVoronoi} compares our method with~\cite{10845125}, which computes the medial axis by segmenting CAD models into patches and computing generalized Voronoi diagrams for each patch. For models amenable to clean segmentation, \cite{10845125} produces reasonable results, though with poor triangle quality. However, the method struggles or fails entirely when faced with models that resist straightforward patch decomposition.

Our method addresses these limitations through its surface-guided framework. The explicit correspondence between medial spheres and surface regions enables accurate feature alignment for sharp-featured models, while the continuous SQEM formulation provides the geometric fidelity needed to capture filleted edges. The result is a unified approach that handles both categories effectively without requiring specialized preprocessing or GPU acceleration.
\begin{figure*}[h]
	\centering
\includegraphics
[width=0.99\linewidth]{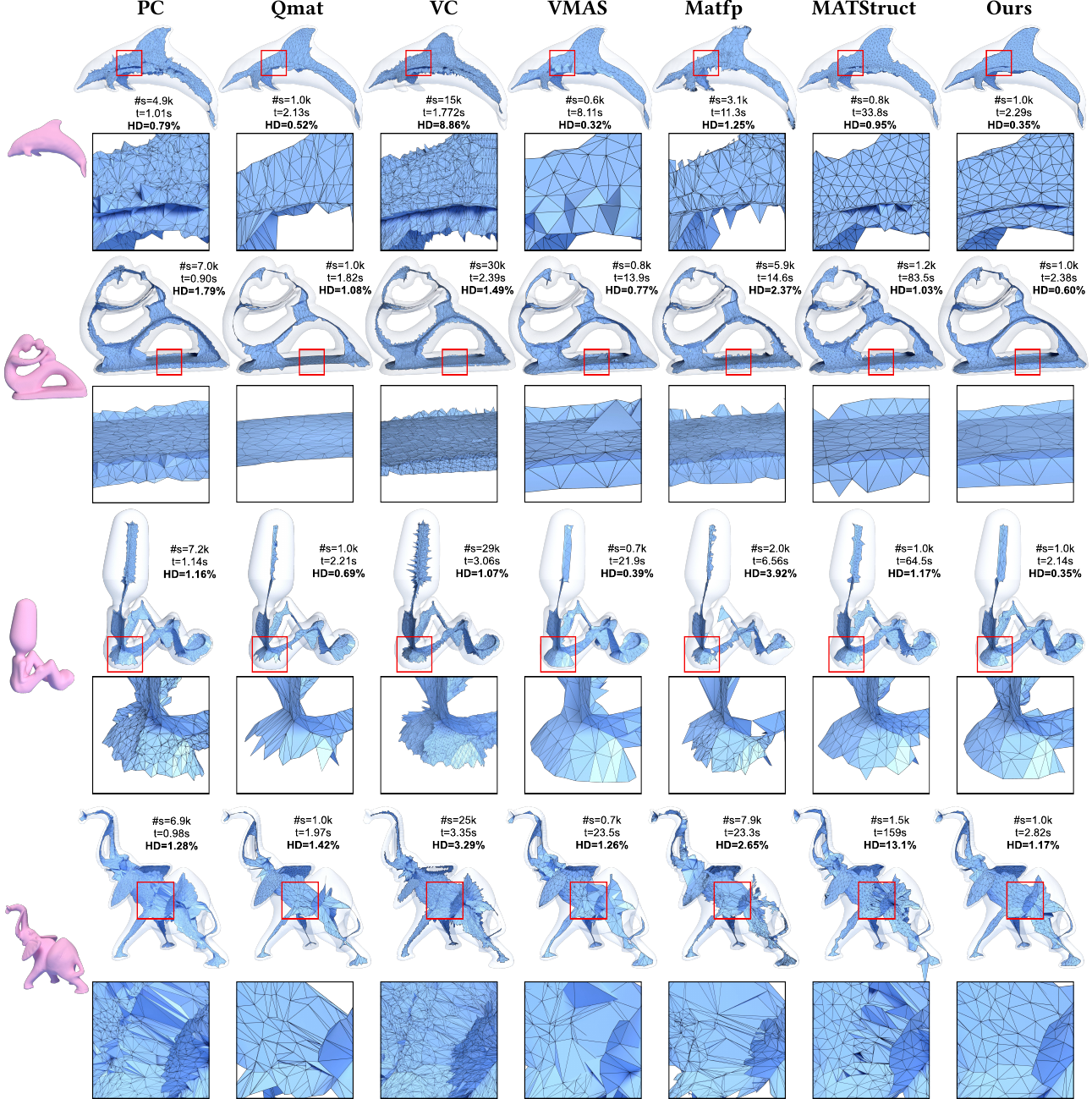}
\caption{Medial axis comparisons on four organic models, showing computation time (t), sphere count (\#s), and the Hausdorff distance (HD) between the reconstructed surface and the original input. }
\label{fig:OrganicComparasion}
\end{figure*}
\subsection{Comparisons on Organic Models}
We further evaluate our method against six competing approaches on multiple organic (freeform) models. Figure~\ref{fig:OrganicComparasion} shows comparative results on four representative examples, annotating computation time ($t$), bidirectional Hausdorff distance (HD), and the number of medial spheres ($\#s$) for each method.

PC generates medial axes containing spurious spike structures (see the first model). Q-MAT employs iterative simplification where each step relies on the previous iteration's result, causing progressive error accumulation that weakens surface fidelity and produces irregular, zigzag boundaries (observe the third model). VC loses important structural details during computation, such as failing to preserve the dolphin's fin in the first model.
VMAS demonstrates strong approximation quality with few spheres through iterative insertion and optimization. However, the method stalls as sphere insertion cannot proceed effectively at higher sphere counts. Additionally, VMAS establishes mesh connectivity based on the surface tessellation induced by medial spheres, which leads to structurally disordered results (models 1, 2, and 4).
MATFP produces results with unwanted spike artifacts. 
MATStruct achieves high triangle quality but produces structurally incorrect results in some cases (models 1 and 4).

Our method demonstrates consistent results across all test cases, maintaining geometric fidelity and mesh quality while avoiding many of the artifacts observed in competing approaches.
\begin{table}[]
\caption{Timing statistics for our method on representative models. For each model, we report the total computation time (in millisecond) and the time distribution across key stages: sampling, RVD computation, 3D Voronoi computation, queue initialization, and edge collapse simplification.}
\label{table:TimeData}
\resizebox{0.99\columnwidth}{!}{%
\begin{tabular}{c|c|cccccc}
\toprule
\multirow{2}{*}{Models}     & \multirow{2}{*}{\# Samples} & \multicolumn{6}{c}{Time (ms)}                                            \\ \cline{3-8}
                            &                             & Sampling & RVD & Voro3D & Init   & Simp   & \multicolumn{1}{l}{Total} \\ 
                            \midrule
 \includegraphics[width=0.6cm, valign=c]{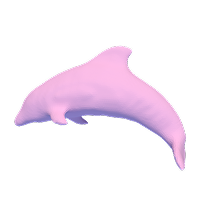}                      & 10K                         & 296.15   & 190.89 & 123.80 & 345.32 & 1337.9 & 2294.0                    \\
 \includegraphics[width=0.6cm, valign=c]{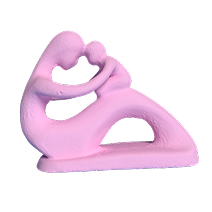}                      & 10K                         & 300.04   & 169.09 & 132.08 & 305.30 & 1470.8 & 2386.1                    \\
 \includegraphics[width=0.6cm, valign=c]{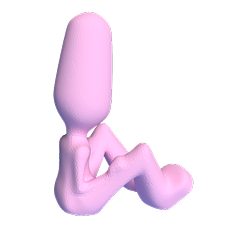}                       & 10K                         & 295.95   & 116.09 & 139.34 & 315.03 & 1277.3 & 2143.7                    \\
 \includegraphics[width=0.6cm, valign=c]{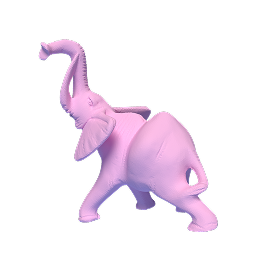}                       & 10K                         & 308.57   & 196.71 & 145.19 & 332.55 & 1822.3 & 2827.2                    \\ \midrule
 \includegraphics[width=0.6cm, valign=c]{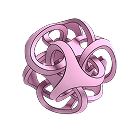}  & 50K                         & 1737.5   & 724.93 & 1139.4 & 1331.7 & 6478.8 & 11444                     \\
 \includegraphics[width=0.6cm, valign=c]{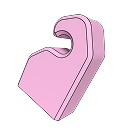}                       & 50K                         & 1301.7   & 359.78 & 939.02 & 864.19 & 6680.3 & 10146                     \\
 \includegraphics[width=0.6cm, valign=c]{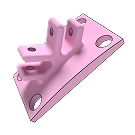}  & 50K                         & 1533.3   & 445.71 & 832.24 & 970.97 & 6640.6 & 10435                     \\
\includegraphics[width=0.6cm, valign=c]{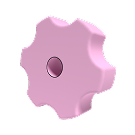}  & 50K                         & 1404.6   & 621.30 & 824.17 & 1040.0 & 6781.9 & 10679                     \\ \bottomrule
\end{tabular}
}
\end{table}
\begin{figure}[h]
	\centering
\includegraphics[width=0.99\linewidth]{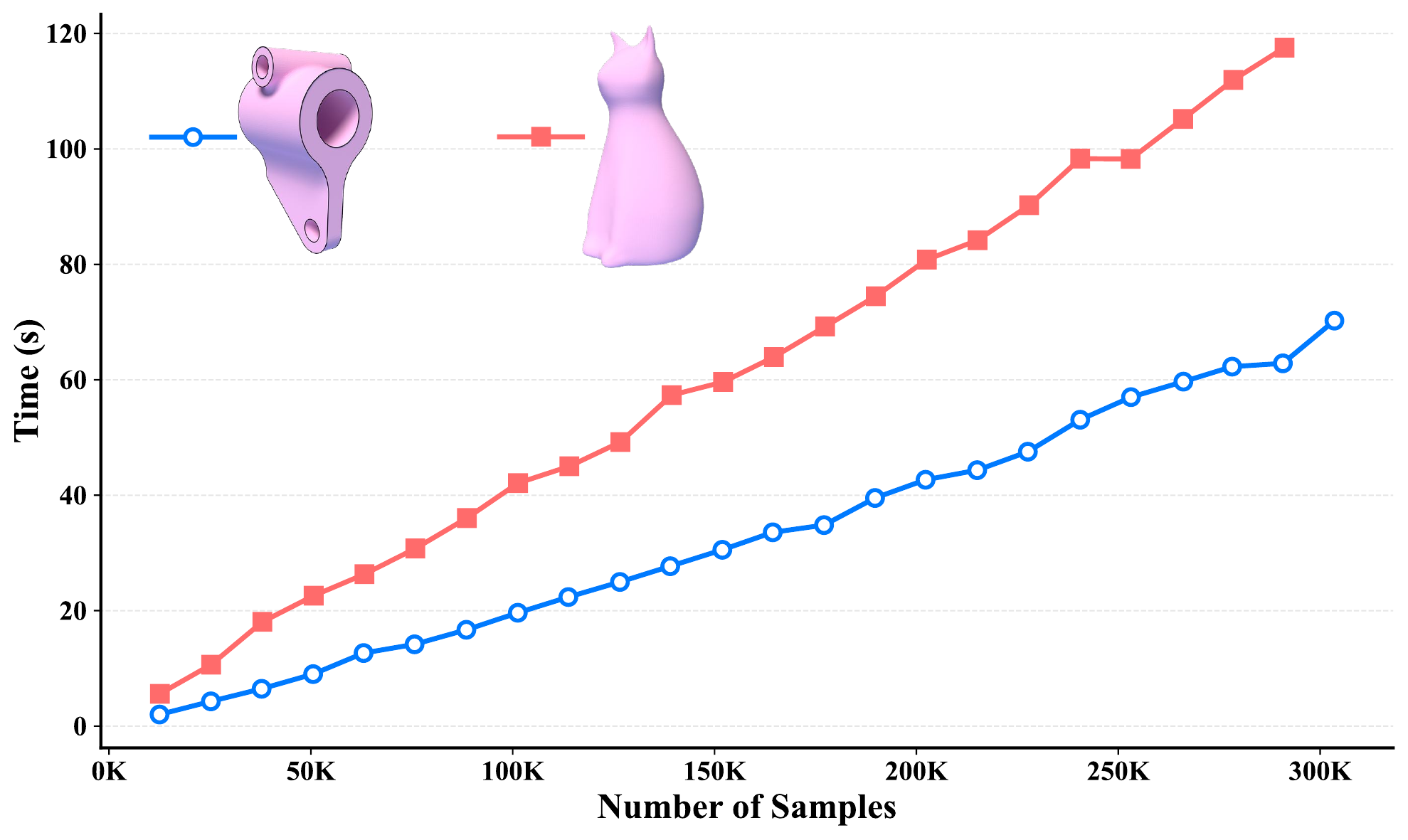}
\caption{Runtime scaling with increasing sample count (fixed simplification target of 1,000 vertices). The timing varies slightly across different models depending on their geometric complexity.}
\label{fig:timeOfSamplesVary}
\end{figure}

\subsection{Performance Analysis}
Our algorithm consists of several key stages: blue noise sampling, RVD computation, 3D Voronoi computation, priority queue initialization, and iterative edge collapse. Benefiting from the additivity property of SQEM matrices, our method achieves efficient computation times.

Table~\ref{table:TimeData} presents the runtime breakdown for each stage on the models from Figures~\ref{fig:CADComparasion} and~\ref{fig:OrganicComparasion}. The results show that edge simplification dominates the overall computation time, accounting for the largest proportion of the total runtime. This is expected, as the iterative collapse process involves repeated energy evaluations and priority queue updates.

Figure~\ref{fig:timeOfSamplesVary} illustrates how computation time scales with the number of input samples (with the medial axis consistently simplified to $1K$ vertices). The timing varies slightly across different models depending on their geometric complexity. For typical models, $50K$ sample points provide sufficient coverage, resulting in total computation times under $15$ seconds. This demonstrates the practical efficiency of our approach—the surface-guided simplification avoids the expensive iterative reconstruction while maintaining high-quality results.

\begin{figure}[h]
	\centering
\includegraphics
[width=0.99\linewidth]{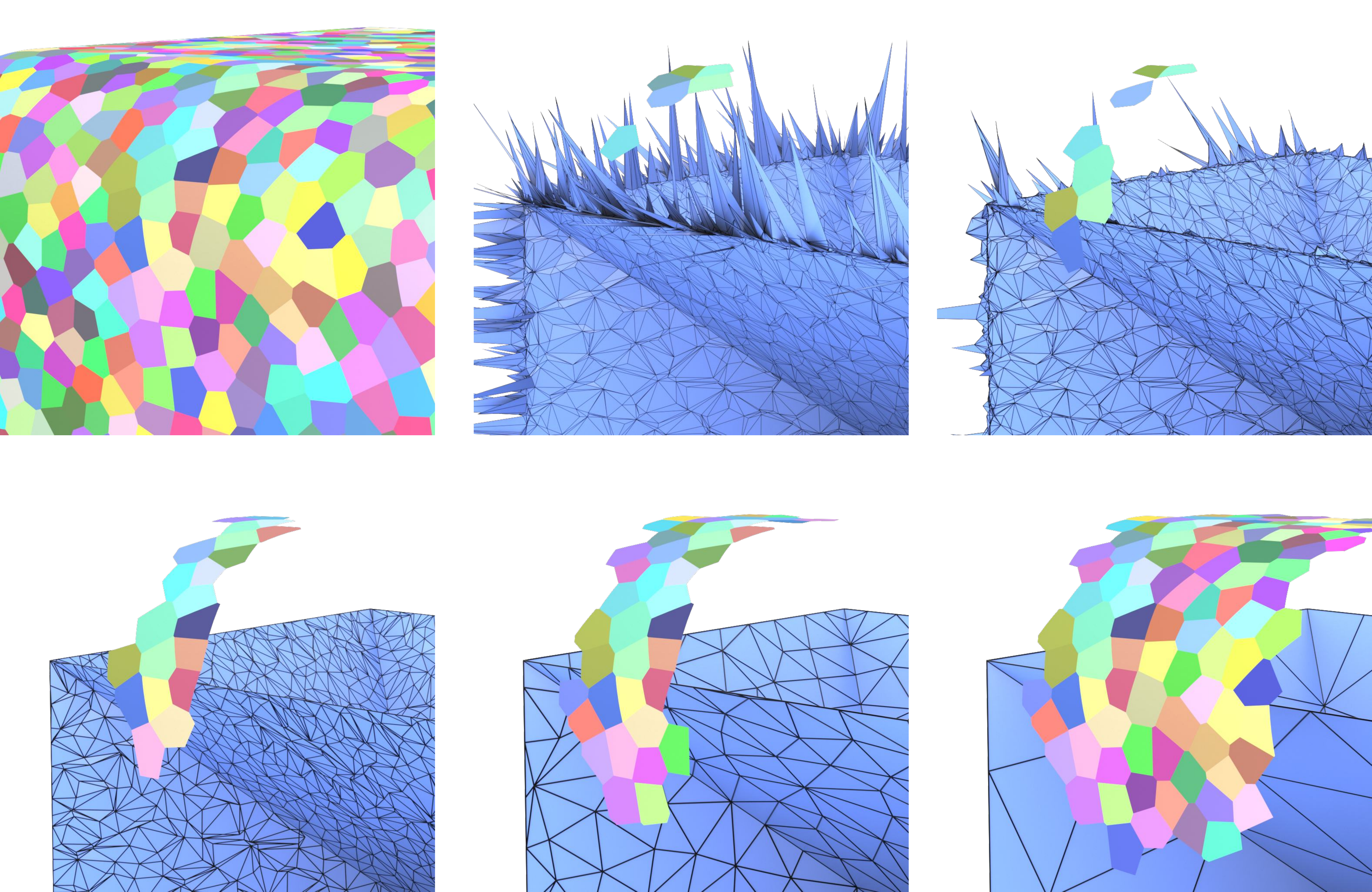}
\caption{Visualization of Atlas evolution across successive simplification stages on a filleted cube model. The top-left subfigure shows the local Restricted Voronoi Diagram (RVD) on the surface. The remaining five subfigures illustrate the surface regions associated with a medial mesh edge at different simplification stages, demonstrating how the Atlas progressively aggregates surface patches as edge collapses proceed.}
\label{fig:Atlases_evolution}
\end{figure}

\begin{figure}[h]
	\centering
\includegraphics
[width=0.99\linewidth]{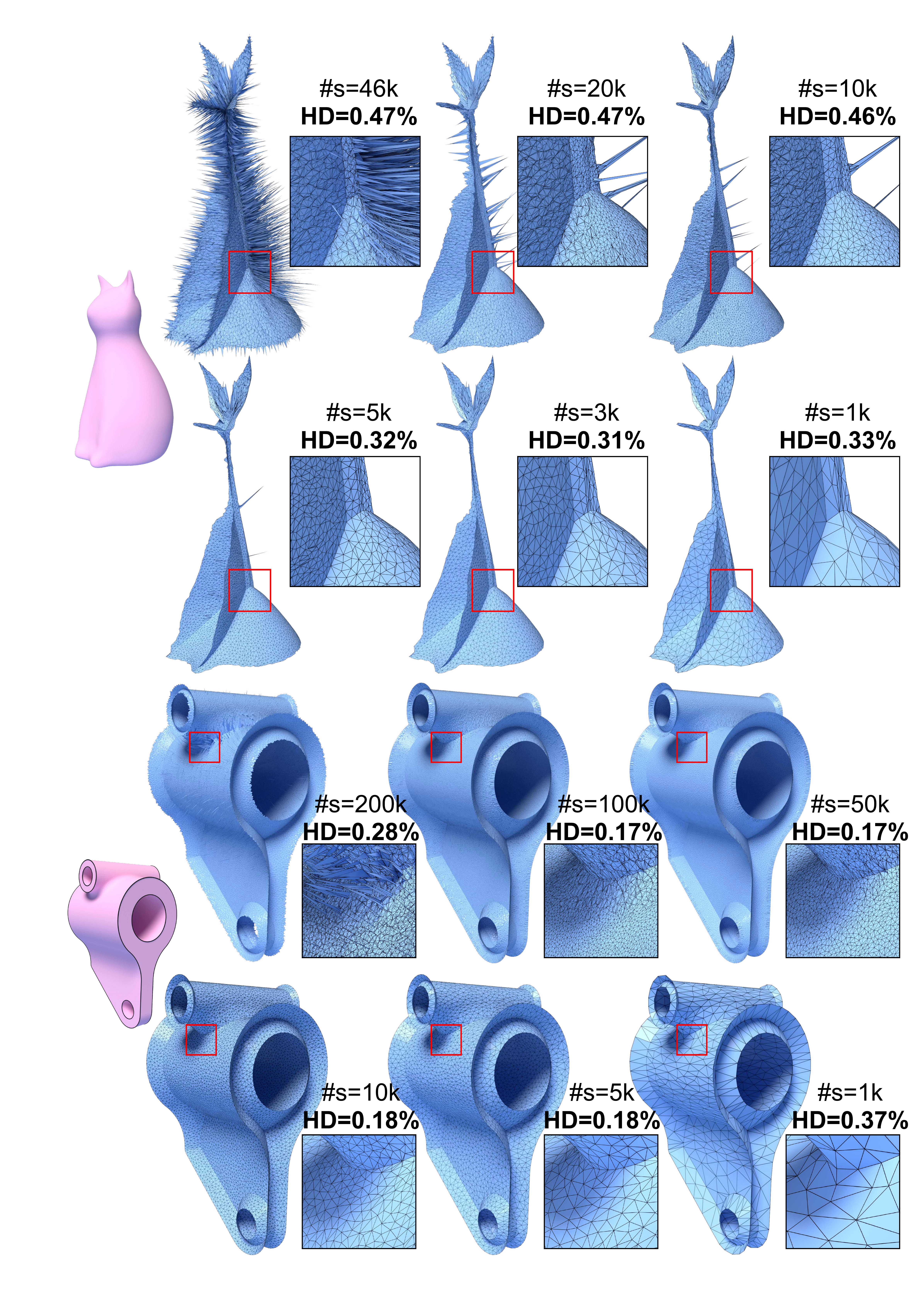}
\caption{Progressive simplification on a CAD model and an organic model. The initial medial axis contains numerous spurious spikes. As simplification progresses, these spikes are removed and triangle quality improves. Notably, the Hausdorff distance (HD) initially decreases before eventually increasing, demonstrating that our surface-guided simplification actively corrects the initial Voronoi structure—intermediate results achieve better approximation than the initialization. In the late stages, as the sphere count becomes very small, approximation capacity naturally declines.}
\label{fig:progressive_simplify}
\end{figure}

\begin{figure}[h]
	\centering
\includegraphics
[width=0.99\linewidth]{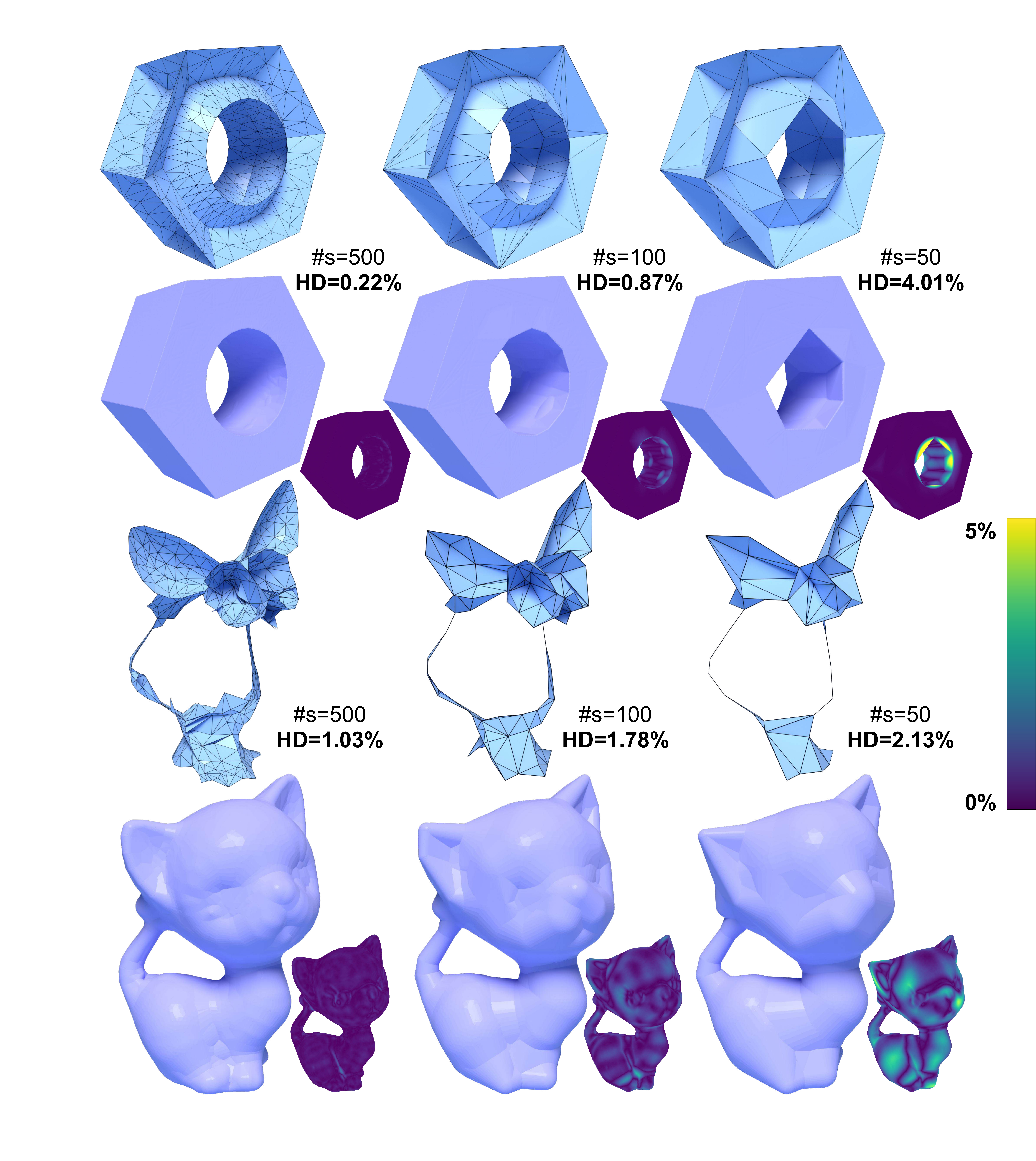}
\caption{Medial axis results simplified to low vertex counts. Despite the sparse sphere count, our method maintains meaningful structural correspondence and produces structurally valid representations even at aggressive simplification levels, demonstrating the effectiveness of the surface-guided optimization strategy.}
\label{fig:progressive_simplify_small_spheres}
\end{figure}
\subsection{Progressive Simplification}
Our method begins with surface sampling and computes the 3D Voronoi diagram as the initial medial axis. Guided by the RVD correspondence, the algorithm progressively simplifies the medial structure through iterative edge collapses until reaching the target number of vertices. 
Figure~\ref{fig:Atlases_evolution} illustrates how the Atlas evolves across simplification stages on a filleted cube model. As simplification proceeds, the medial structure progressively converges toward the centerline of the fillet region, with the associated Atlas accurately capturing the corresponding surface patches.

Figure~\ref{fig:progressive_simplify} illustrates this process on both a CAD model and an organic model, showing intermediate results from the initial medial axis to the final simplified structure, along with the bidirectional Hausdorff distance (HD) and sphere count at each stage.
The initial medial axis contains numerous unwanted spike structures, exhibits poor triangle quality, and fails to accurately capture geometric features. As simplification progresses, these spurious spikes gradually disappear, triangle quality improves, and feature alignment becomes increasingly evident. Notably, the HD metric initially decreases before eventually increasing again, revealing a fundamental characteristic of our approach: rather than progressively refining an inherited medial structure, our method simplifies directly with respect to the original surface geometry. The intermediate simplified results actually achieve better surface approximation than the initial Voronoi-based medial axis, demonstrating that our surface-guided framework actively improves geometric fidelity during simplification. Only in the late stages, when the sphere count becomes very small, does approximation capacity naturally decline.

Figure~\ref{fig:progressive_simplify_small_spheres} further demonstrates the approximation capability of our method when simplified to very few vertices on both CAD and organic models. 
Even with aggressive simplification, the medial axis retains meaningful geometric representation.
To more clearly illustrate the reconstruction fidelity, we additionally provide a per-point Hausdorff distance visualization using a color-coded scalar field, where the color at each surface point reflects its distance to the reconstructed surface. The visualization confirms that reconstruction error remains well-distributed across the surface, with larger deviations concentrated in geometrically complex regions as the sphere count decreases.

\begin{figure}[h]
	\centering
\includegraphics
[width=0.99\linewidth]{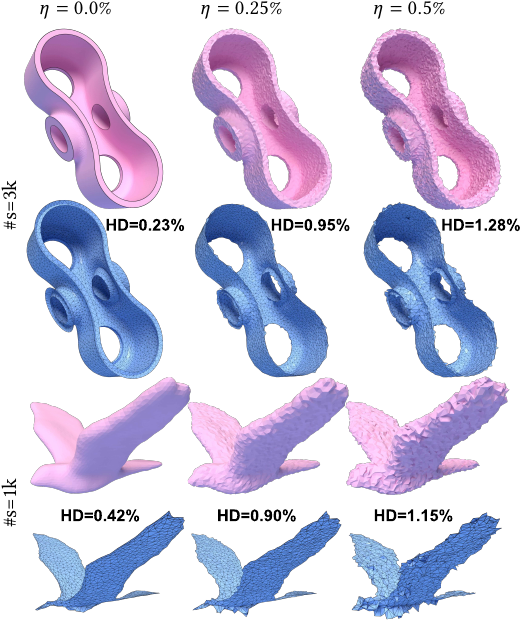}
\caption{Robustness to noise. We add random displacements with different magnitudes ($\eta = 0, 0.25\%, 0.5\%$) to test models. Noise introduces normal discontinuities that activate our Fidelity term for invaginated RVD cells, enabling our method to produce reasonable medial structures across the tested noise levels.}
\label{fig:noise}
\end{figure}

\subsection{Robustness to Noise}
We further evaluate the robustness of our method on noisy inputs. Following standard practice, we add random displacements to each vertex of the input models (one CAD model and one organic model). The displacement magnitude for each vertex is randomly sampled from $[0, \eta \cdot d]$, where $d$ is the diagonal length of the bounding box and $\eta$ controls the maximum noise level. Figure~\ref{fig:noise} shows our results under different noise levels: no noise, $\eta = 0.25\%$, and $\eta = 0.5\%$.

When models contain noise, surface normals become unreliable—the local tangent planes defined by sample points and their normals no longer accurately represent the underlying geometry, rendering the plane-based SQEM energy inaccurate. 
However, our method demonstrates noise resilience.
Noise introduces local normal discontinuities that naturally trigger our \textit{invaginated cell} detection, activating the Fidelity term for invaginated RVD cells. Since this formulation directly measures distances to sample sites rather than relying on tangent plane approximations, it provides more robust optimization under noisy conditions.
As shown in Figure~\ref{fig:noise}, our method produces reasonable medial structures across the tested noise levels.

\begin{figure}[htbp]
	\centering
\includegraphics
[width=0.99\linewidth]{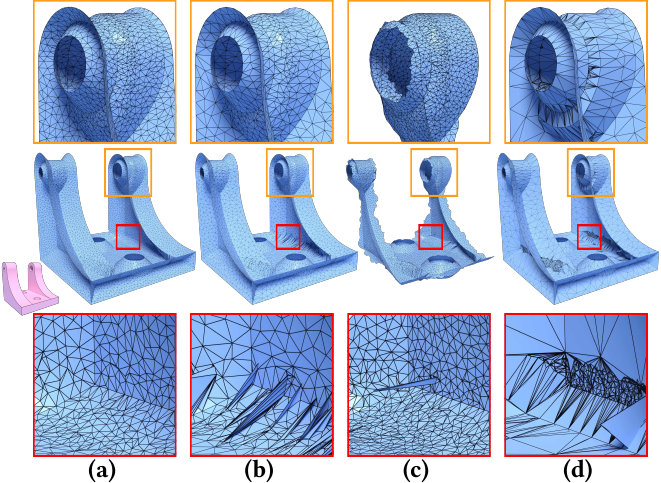}
\caption{Ablation study on a CAD model. (a) Full method. (b) Without Medial Face Filtering near Concave Features. (c) Without sharp feature preservation—external boundaries collapse inward, losing sharp feature alignment. (d) Without Fidelity term for invaginated RVD cells—spheres cluster excessively and produce incorrect medial structures.}
\label{fig:ablation_study}
\end{figure}
\subsection{Ablation Studies and Parameter Analysis}

\textbf{Component Ablation.} We evaluate the contribution of key components in our method through ablation studies. Figure~\ref{fig:ablation_study} shows results on a CAD model under four configurations: the full method, without Medial Face Filtering near Concave Features, without sharp feature preservation, and without Fidelity term for invaginated RVD cells.

Without Medial Face Filtering near Concave Features results in spurious complex structures in concave regions, as edge collapse alone cannot effectively remove non-medial faces in these areas. Without sharp feature preservation, external boundaries progressively collapse inward, failing to preserve the sharp feature information characteristic of CAD models. Disabling Fidelity term for invaginated RVD cells causes spheres to cluster excessively and produces incorrect medial structures in these areas. These ablations confirm that each component plays a critical role in achieving accurate results.

\begin{figure}[htbp]
	\centering
\includegraphics[width=1.0\linewidth]{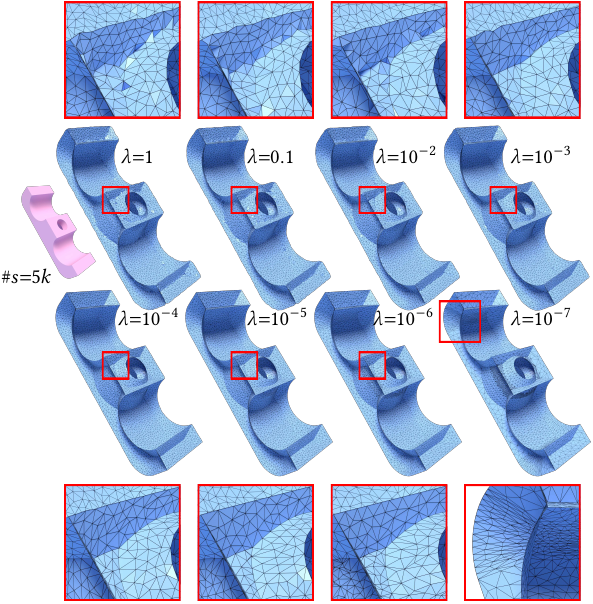}
\caption{Effect of the weight parameter $\lambda$ on medial axis results. Larger $\lambda$ values prioritize mesh quality but weaken feature capture—internal and external features become less distinct. As $\lambda$ decreases, feature alignment becomes increasingly prominent. When $\lambda$ is too small, mesh quality degrades as the method allocates excessive triangles to geometrically complex regions.}
\label{fig:w_paramter_vary}
\end{figure}

\begin{figure}[htbp]
	\centering
\includegraphics
[width=1.0\linewidth]{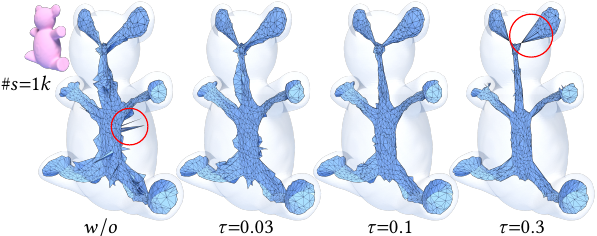}
\caption{Effect of the stability threshold parameter $\tau$ on spike removal. The parameter $\tau$ controls the strictness of spike classification. Without spike filtering, spurious branches cannot be reliably removed. As $\tau$ increases, spike removal becomes more aggressive. However, when $\tau$ is too large, legitimate medial structures are incorrectly removed, weakening approximation capability.}
\label{fig:tau_paramter_study}
\end{figure}

\textbf{Parameter Sensitivity.} We also examine the effect of two key parameters. The weight $\lambda$ balances surface fidelity against mesh quality. Figure~\ref{fig:w_paramter_vary} shows results as $\lambda$ varies from $10^{-7}$ to $1$. Larger $\lambda$ values prioritize mesh quality but reduce approximation fidelity, failing to accurately capture internal and external features. As $\lambda$ decreases, features become increasingly prominent. When $\lambda$ is too small, the method allocates excessive triangles to geometrically complex regions, degrading mesh quality.

The threshold $\tau$ determines the strictness of spike classification. Figure~\ref{fig:tau_paramter_study} shows results across different $\tau$ values. Without spike filtering, standard simplification criteria fail to remove spurious branches effectively. As $\tau$ increases, spike removal becomes more aggressive. However, when $\tau$ is too large, the method removes legitimate medial structures, weakening approximation capability.

Despite these variations, our method produces acceptable results across a reasonable parameter range, demonstrating robustness to parameter choices.

\paragraph{Fixed Parameters.} 
Beyond $\lambda$ and $\tau$, our method involves several additional parameters that are fixed throughout all experiments. The dihedral angle tolerance $\phi$ controls the detection of sharp and concave features: we set $\phi = 45^\circ$ in all experiments, which provides reliable feature detection across a wide range of CAD models without requiring per-model tuning. The filtering threshold $\alpha = 0.7$ in the medial face filtering step (Equation~\ref{eq:concave:filter}) is also kept constant; values in the range $[0.6, 0.8]$ yield consistent results, and the method is not sensitive to moderate variations in this parameter. 
The topology preservation threshold is set to 200 vertices, below which the Link Condition is enforced prior to every edge collapse; this value is chosen to reduce the risk of undesired topological changes during simplification, and can be increased for models with more complex topology. 
The sharpness parameter $k = 100$ in the sigmoid weighting function $\Psi$ enforces a near-binary transition between spike removal and geometry-guided simplification; in practice, any sufficiently large value (e.g., $k \geq 50$) produces equivalent behavior. The number of surface samples is set to $10K$ for organic models and $50K$ for CAD models, which we find to provide sufficient geometric coverage for typical inputs.

\begin{figure}[htbp]
	\centering
\includegraphics
[width=0.99\linewidth]{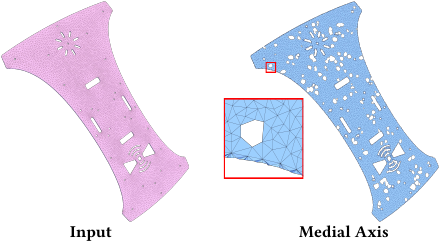}
\caption{A representative failure case on a thin sheet-like model. With 15K surface samples, the insufficient sampling density causes the 3D Voronoi diagram to penetrate through the thin surface during initialization, resulting in holes in the medial structure, and also prevents accurate capture of the side features.}
\label{fig:failure}
\end{figure}

Despite its general effectiveness, our method does encounter difficulties in certain challenging configurations. Figure~\ref{fig:failure} shows a representative failure case on a thin sheet-like model: insufficient sampling density on such thin surfaces causes the 3D Voronoi diagram to penetrate through the surface, resulting in holes in the medial structure, and also fails to capture side features accurately.

\subsection{Potential Application}
\paragraph{Fillet Detection}
Filleting is a fundamental operation in CAD systems, creating smooth transitions between adjacent surface patches—akin to a ball rolling between two surfaces to form a seamless connection. While fillets appear as narrow transitional regions on the surface, they play a crucial role in reverse engineering and secondary design phases. However, fillet detection remains challenging, particularly when input data originates from surface reconstruction or discretization processes~\cite{10.1145/3731166}.

\begin{figure}[htbp]
	\centering
\includegraphics
[width=0.99\linewidth]{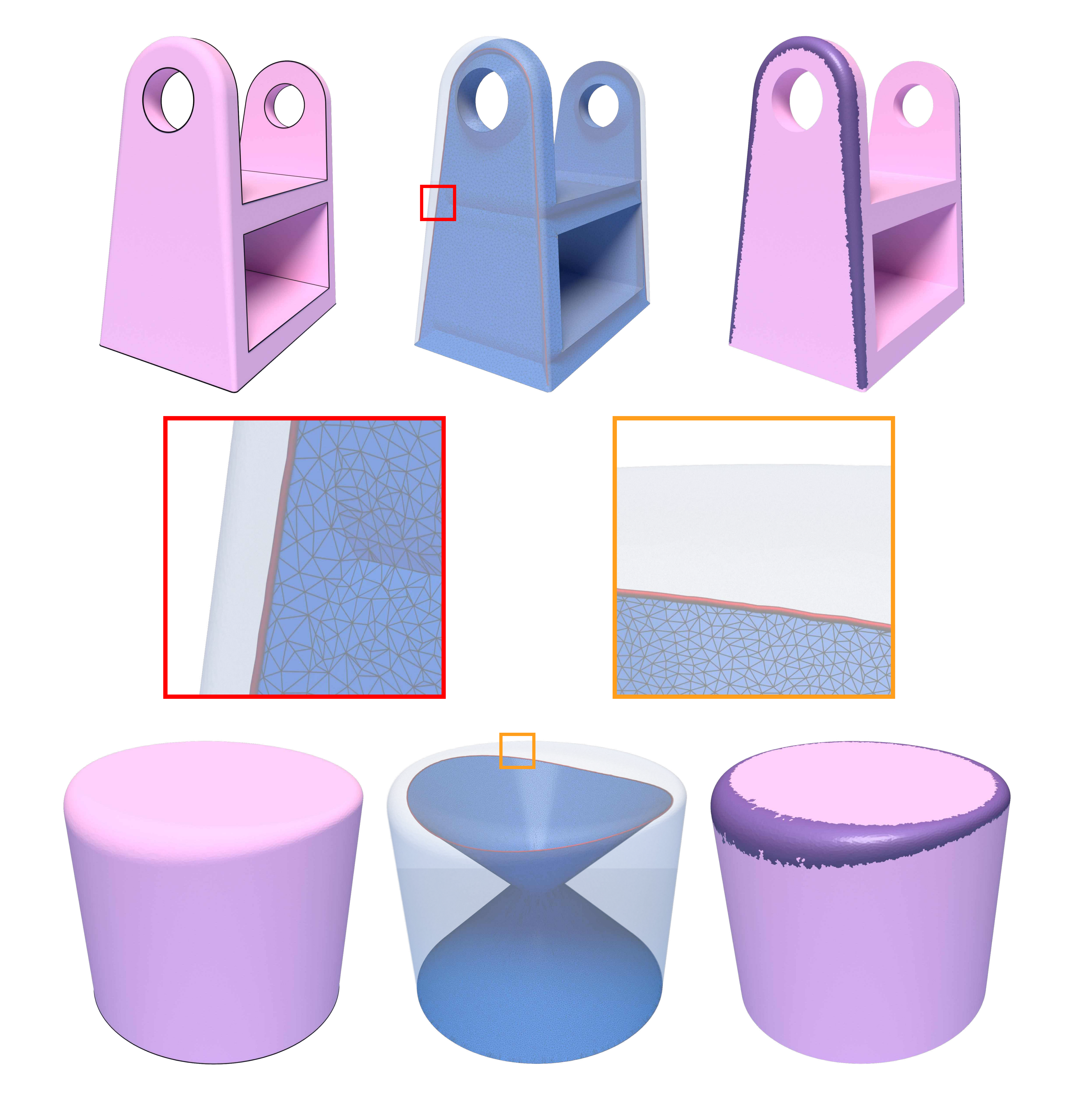}
\caption{Fillet detection application. Our surface-guided framework accurately captures fillet centerlines (red), including variable-radius cases. The natural encoding of radius values and surface correspondence in medial edges enables straightforward fillet detection through distance consistency verification.}
\label{fig:roundEdge}
\end{figure}

The medial axis provides a natural geometric lens for understanding fillets. While medial axes generally consist of both variable-radius and constant-radius branches, the medial structure of a fillet converges to the centerline traced by the rolling sphere centers—a well-defined curve with smoothly varying radius. Benefiting from our method's accurate geometric capture capability, the computed medial axis precisely identifies these fillet centerlines. Figure~\ref{fig:roundEdge} shows our results on two models containing variable-radius fillets, where the red-highlighted edges accurately trace the fillet centerlines. Our medial representation naturally encodes the information needed for fillet detection: each medial edge carries both radius values (from optimization) and explicit correspondence to surface regions (through the dual Voronoi framework). This enables a straightforward detection strategy—starting from medial boundary edges, we verify whether the Euclidean distances from corresponding surface triangles to the medial edge consistently match the optimized radius values. Edges satisfying this criterion are classified as fillet centerlines, as demonstrated in Figure~\ref{fig:roundEdge}. These results highlight the potential of our surface-guided framework for this important CAD analysis task.

\paragraph{Surface Extraction from Unsigned Distance Fields}
Unsigned Distance Fields (UDFs) are a widely adopted shape representation in computer graphics and vision, supporting a broad range of applications including neural implicit modeling and shape reconstruction. A fundamental challenge in UDF-based pipelines is surface extraction: unlike signed distance fields, UDFs lack sign information and therefore cannot be directly processed by standard approaches such as Marching Cubes. One prominent line of work~\cite{Hou2023DCUDF,11079233} addresses this by first extracting an $\epsilon$-isosurface, yielding a double-layered structure that is subsequently projected toward the zero level set via optimization, followed by a post-processing step to recover a single-layer surface. However, the optimization is sensitive to parameter choices and prone to geometric artifacts in high-curvature regions, while the double-to-single-layer transition remains fragile and can fail on non-manifold or non-orientable inputs.

In this context, the medial axis of the double-layered $\epsilon$-isosurface naturally corresponds to the underlying surface encoded by the UDF. This medial representation inherently supports non-manifold structures and directly yields a single-layer output, bypassing the need for explicit collapse optimization. We believe that a high-quality medial axis computation can serve as a principled and robust foundation for this extraction step.

\begin{figure}[htbp]
	\centering
\includegraphics
[width=0.99\linewidth]{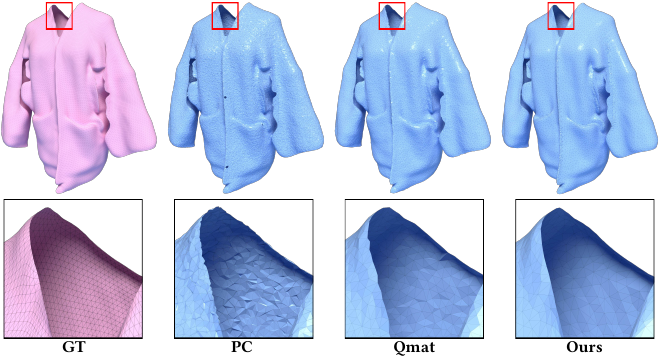}
\caption{Surface extraction from an unsigned distance field on a garment model from the Deep Fashion3D V2 dataset~\cite{zhu2020deep}, comparing the ground truth surface against extraction results using three representative medial axis methods. Our result exhibits cleaner boundaries and closer geometric alignment to the ground truth, demonstrating the potential of our surface-guided framework for UDF-based surface extraction.}
\label{fig:udf_extraction}
\end{figure}

To demonstrate this potential, we take a garment model from the Deep Fashion3D V2 dataset~\cite{zhu2020deep} as a representative example. We extract the $\epsilon$-isosurface near the surface and apply several representative medial axis methods for comparison. As shown in Figure~\ref{fig:udf_extraction}, our method produces a cleaner result with more regular boundaries and closer alignment to the ground truth surface, highlighting the promise of our surface-guided framework for UDF-based surface extraction.

\section{Limitations and Future Work}

Our method has several limitations that suggest directions for future research. First, the 3D Voronoi-based initialization faces challenges with extremely thin sheet-like models, where insufficient thickness may prevent the formation of a valid initial medial structure. Additionally, the simplified medial mesh may exhibit slight self-intersections, particularly in regions with complex branching or high curvature.
Finally, our current topology preservation strategy is heuristic in nature and does not provide formal guarantees; topological errors may still occur, particularly on models with high topological complexity.

Looking forward, we plan to investigate more robust initialization strategies that can handle extremely thin geometric features, potentially through adaptive sampling or alternative initialization schemes.
We also intend to explore more principled topology preservation mechanisms that can provide stronger guarantees during simplification. 
Addressing these challenges would further enhance the method's applicability to a broader range of geometric modeling and analysis tasks.

\section{Conclusion}
We have presented Structural MAT, a robust framework for medial axis simplification that maintains explicit correspondence between medial axis vertices and surface regions throughout the entire simplification process. By tracking the association between each 3D medial element and its corresponding surface patch—what we term the Atlas—our method achieves clean boundaries, accurate feature preservation, and high computational scalability. Our approach naturally handles challenging geometric features including sharp edges and filleted surfaces, which have traditionally been difficult for medial axis methods, making it particularly effective for both organic shapes and precision-engineered CAD models.

Extensive evaluation demonstrates that Structural MAT consistently outperforms existing methods across diverse geometric datasets, achieving superior shape fidelity while maintaining computational efficiency. The demonstrated application in fillet detection showcases the utility of our surface-aware representation for high-level geometric reasoning tasks, confirming that explicit surface correspondence serves as a powerful foundation for downstream applications in computer-aided design and manufacturing.

\begin{acks}
    The authors thank the anonymous reviewers for their insightful comments and suggestions. This work was supported by the National Natural Science Foundation of China (Key Project Grants 12494550 and 12494553; Grants U23A20312 and 62272277) and the Natural Science Foundation of Shandong Province (Grant ZR2025MS986).
\end{acks}

\bibliographystyle{ACM-Reference-Format}
\bibliography{main}

@String{Computing = "Computing" }

@String{Computer = "{IEEE} Computer" }

@String{Springer = "Springer-Verlag" }

@ArtifactSoftware{R,
    title = {R: A Language and Environment for Statistical Computing},
    author = {{R Core Team}},
    organization = {R Foundation for Statistical Computing},
    address = {Vienna, Austria},
    year = {2019},
    url = {https://www.R-project.org/},
}

@InProceedings{Koch_2019_CVPR,
author = {Koch, Sebastian and Matveev, Albert and Jiang, Zhongshi and Williams, Francis and Artemov, Alexey and Burnaev, Evgeny and Alexa, Marc and Zorin, Denis and Panozzo, Daniele},
title = {ABC: A Big CAD Model Dataset For Geometric Deep Learning},
booktitle = {The IEEE Conference on Computer Vision and Pattern Recognition (CVPR)},
month = {June},
year = {2019}
}

@article{Thingi10K,
  title={Thingi10K: A Dataset of 10,000 3D-Printing Models},
  author={Zhou, Qingnan and Jacobson, Alec},
  journal={arXiv preprint arXiv:1605.04797},
  year={2016}
}

@article{wang2022matfp,
    title={Computing medial axis transform with feature preservation via restricted power diagram},
    author={Wang, Ningna and Wang, Bin and Wang, Wenping and Guo, Xiaohu},
    journal={ACM Transactions on Graphics (Proc. SIGGRAPH Asia)},
    volume={41},
    number={6},
    pages={1--18},
    year={2022},
    publisher={ACM New York, NY, USA}
}

@software{FCPW,
author = {Sawhney, Rohan},
title = {FCPW: Fastest Closest Points in the West},
version = {1.0},
year = {2021}
}

@ARTICLE{11165079,
author={Wang, Pengfei and Song, Jiantao and Xin, Shiqing and Chen, Shuangmin and Tu, Changhe and Wang, Wenping and Wang, Jiaye},
journal={IEEE Transactions on Pattern Analysis and Machine Intelligence},
title={Efficient Nearest Neighbor Search Using Dynamic Programming},
year={2025},
volume={},
number={},
pages={1-16},
doi={10.1109/TPAMI.2025.3610211}
}

@book{siddiqi2008medial,
  title={Medial representations: mathematics, algorithms and applications},
  author={Siddiqi, Kaleem and Pizer, Stephen},
  volume={37},
  year={2008},
  publisher={Springer Science \& Business Media}
}

@inproceedings{tagliasacchi20163d,
  title={3d skeletons: A state-of-the-art report},
  author={Tagliasacchi, Andrea and Delame, Thomas and Spagnuolo, Michela and Amenta, Nina and Telea, Alexandru},
  booktitle={Computer Graphics Forum},
  volume={35},
  number={2},
  pages={573--597},
  year={2016},
  organization={Wiley Online Library}
}

@article{brandt1994convergence,
  title={Convergence and continuity criteria for discrete approximations of the continuous planar skeleton},
  author={Brandt, Jonathan W},
  journal={CVGIP: Image Understanding},
  volume={59},
  number={1},
  pages={116--124},
  year={1994},
  publisher={Elsevier}
}

@article{amenta2001power,
  title={The power crust, unions of balls, and the medial axis transform},
  author={Amenta, Nina and Choi, Sunghee and Kolluri, Ravi Krishna},
  journal={Computational Geometry},
  volume={19},
  number={2-3},
  pages={127--153},
  year={2001},
  publisher={Elsevier}
}

@article{dey2004approximating,
  title={Approximating the medial axis from the Voronoi diagram with a convergence guarantee},
  author={Dey, Tamal K and Zhao, Wulue},
  journal={Algorithmica},
  volume={38},
  number={1},
  pages={179--200},
  year={2004},
  publisher={Springer}
}

@inproceedings{attali1996modeling,
  title={Modeling noise for a better simplification of skeletons},
  author={Attali, Dominique and Montanvert, Annick},
  booktitle={Proceedings of 3rd IEEE International Conference on Image Processing},
  volume={3},
  pages={13--16},
  year={1996},
  organization={IEEE}
}

@article{chazal2008smooth,
  title={Smooth manifold reconstruction from noisy and non-uniform approximation with guarantees},
  author={Chazal, Fr{\'e}d{\'e}ric and Lieutier, Andr{\'e}},
  journal={Computational Geometry},
  volume={40},
  number={2},
  pages={156--170},
  year={2008},
  publisher={Elsevier}
}

@article{chazal2005lambda,
  title={The “$\lambda$-medial axis”},
  author={Chazal, Fr{\'e}d{\'e}ric and Lieutier, Andr{\'e}},
  journal={Graphical models},
  volume={67},
  number={4},
  pages={304--331},
  year={2005},
  publisher={Elsevier}
}

@inproceedings{giesen2009scale,
  title={The scale axis transform},
  author={Giesen, Joachim and Miklos, Balint and Pauly, Mark and Wormser, Camille},
  booktitle={Proceedings of the twenty-fifth annual symposium on Computational geometry},
  pages={106--115},
  year={2009}
}

@incollection{miklos2010discrete,
  title={Discrete scale axis representations for 3D geometry},
  author={Miklos, Balint and Giesen, Joachim and Pauly, Mark},
  booktitle={ACM SIGGRAPH 2010 papers},
  pages={1--10},
  year={2010}
}

@article{hesselink2008euclidean,
  title={Euclidean skeletons of digital image and volume data in linear time by the integer medial axis transform},
  author={Hesselink, Wim H and Roerdink, Jos BTM},
  journal={IEEE Transactions on Pattern Analysis and Machine Intelligence},
  volume={30},
  number={12},
  pages={2204--2217},
  year={2008},
  publisher={IEEE}
}

@incollection{rumpf2002continuous,
  title={A continuous skeletonization method based on level sets},
  author={Rumpf, Martin and Telea, Alexandru},
  booktitle={EPRINTS-BOOK-TITLE},
  year={2002},
  publisher={University of Groningen, Johann Bernoulli Institute for Mathematics and~…}
}

@article{pudney1998distance,
  title={Distance-ordered homotopic thinning: a skeletonization algorithm for 3D digital images},
  author={Pudney, Chris},
  journal={Computer vision and image understanding},
  volume={72},
  number={3},
  pages={404--413},
  year={1998},
  publisher={Elsevier}
}

@article{yan2018voxel,
  title={Voxel cores: Efficient, robust, and provably good approximation of 3d medial axes},
  author={Yan, Yajie and Letscher, David and Ju, Tao},
  journal={ACM Transactions on Graphics (TOG)},
  volume={37},
  number={4},
  pages={1--13},
  year={2018},
  publisher={ACM New York, NY, USA}
}

@article{wang2024mattopo,
  title     = {MATTopo: Topology-preserving Medial Axis Transform with Restricted Power Diagram},
  author    = {Wang, Ningna and Huang, Hui and Song, Shibo and Wang, Bin and Wang, Wenping and Guo, Xiaohu},
  journal   = {ACM Transactions on Graphics (TOG)},
  year      = {2024},
  address   = {New York, NY, USA},
  publisher = {ACM},
  volume    = {43},
  number    = {4},
  url       = {https://doi.org/10.1145/3687763},
  doi       = {10.1145/3687763},
  booktitle = {ACM SIGGRAPH Asia 2024 Papers},
  series    = {SIGGRAPH Asia '24}
}

@inproceedings{10.1145/3757377.3763840,
author = {Wang, Ningna and Xu, Rui and Yin, Yibo and Zhong, Zichun and Komura, Taku and Wang, Wenping and Guo, Xiaohu},
title = {MATStruct: High-quality Medial Mesh Computation via Structure-aware Variational Optimization},
year = {2025},
isbn = {9798400721373},
publisher = {Association for Computing Machinery},
address = {New York, NY, USA},
url = {https://doi.org/10.1145/3757377.3763840},
doi = {10.1145/3757377.3763840},
booktitle = {Proceedings of the SIGGRAPH Asia 2025 Conference Papers},
articleno = {170},
numpages = {12},
location = {
},
series = {SA Conference Papers '25}
}

@article{10.1145/2753755,
author = {Li, Pan and Wang, Bin and Sun, Feng and Guo, Xiaohu and Zhang, Caiming and Wang, Wenping},
title = {Q-MAT: Computing Medial Axis Transform By Quadratic Error Minimization},
year = {2016},
issue_date = {December 2015},
publisher = {Association for Computing Machinery},
address = {New York, NY, USA},
volume = {35},
number = {1},
issn = {0730-0301},
url = {https://doi.org/10.1145/2753755},
doi = {10.1145/2753755},
journal = {ACM Trans. Graph.},
month = dec,
articleno = {8},
numpages = {16}
}

@inproceedings{dou2022coverage,
  title={Coverage axis: Inner point selection for 3d shape skeletonization},
  author={Dou, Zhiyang and Lin, Cheng and Xu, Rui and Yang, Lei and Xin, Shiqing and Komura, Taku and Wang, Wenping},
  booktitle={Computer Graphics Forum},
  volume={41},
  number={2},
  pages={419--432},
  year={2022},
  organization={Wiley Online Library}
}

@inproceedings{wang2024coverage,
  title={Coverage axis++: Efficient inner point selection for 3D shape skeletonization},
  author={Wang, Zimeng and Dou, Zhiyang and Xu, Rui and Lin, Cheng and Liu, Yuan and Long, Xiaoxiao and Xin, Shiqing and Komura, Taku and Yuan, Xiaoming and Wang, Wenping},
  booktitle={Computer Graphics Forum},
  volume={43},
  number={5},
  pages={e15143},
  year={2024},
  organization={Wiley Online Library}
}

@inproceedings{10.1145/3680528.3687678,
author = {Huang, Qijia and Kraemer, Pierre and Thery, Sylvain and Bechmann, Dominique},
title = {Dynamic Skeletonization via Variational Medial Axis Sampling},
year = {2024},
isbn = {9798400711312},
publisher = {Association for Computing Machinery},
address = {New York, NY, USA},
url = {https://doi.org/10.1145/3680528.3687678},
doi = {10.1145/3680528.3687678},
booktitle = {SIGGRAPH Asia 2024 Conference Papers},
articleno = {66},
numpages = {11},
location = {Tokyo, Japan},
series = {SA '24}
}

@article{dou2020top,
  title={Top-down shape abstraction based on greedy pole selection},
  author={Dou, Zhiyang and Xin, Shiqing and Xu, Rui and Xu, Jian and Zhou, Yuanfeng and Chen, Shuangmin and Wang, Wenping and Zhao, Xiuyang and Tu, Changhe},
  journal={IEEE transactions on visualization and computer graphics},
  volume={27},
  number={10},
  pages={3982--3993},
  year={2020},
  publisher={IEEE}
}

@article{fu2022easyvrmodeling,
  title={Easyvrmodeling: Easily create 3d models by an immersive vr system},
  author={Fu, Zhiying and Xu, Rui and Xin, Shiqing and Chen, Shuangmin and Tu, Changhe and Yang, Chenglei and Lu, Lin},
  journal={Proceedings of the ACM on Computer Graphics and Interactive Techniques},
  volume={5},
  number={1},
  pages={1--14},
  year={2022},
  publisher={ACM New York, NY, USA}
}

@inproceedings{hu2019mat,
  title={MAT-Net: Medial Axis Transform Network for 3D Object Recognition.},
  author={Hu, Jianwei and Wang, Bin and Qian, Lihui and Pan, Yiling and Guo, Xiaohu and Liu, Lingjie and Wang, Wenping},
  booktitle={IJCAI},
  pages={774--781},
  year={2019}
}

@article{noma2024surface,
  title={Surface-filling curve flows via implicit medial axes},
  author={Noma, Yuta and Sell{\'a}n, Silvia and Sharp, Nicholas and Singh, Karan and Jacobson, Alec},
  journal={ACM Transactions on Graphics (TOG)},
  volume={43},
  number={4},
  pages={1--12},
  year={2024},
  publisher={ACM New York, NY, USA}
}

@article{zhou2015generalized,
  title={Generalized cylinder decomposition.},
  author={Zhou, Yang and Yin, Kangxue and Huang, Hui and Zhang, Hao and Gong, Minglun and Cohen-Or, Daniel},
  journal={ACM Trans. Graph.},
  volume={34},
  number={6},
  pages={171--1},
  year={2015}
}

@article{lin2020seg,
  title={Seg-mat: 3d shape segmentation using medial axis transform},
  author={Lin, Cheng and Liu, Lingjie and Li, Changjian and Kobbelt, Leif and Wang, Bin and Xin, Shiqing and Wang, Wenping},
  journal={IEEE transactions on visualization and computer graphics},
  volume={28},
  number={6},
  pages={2430--2444},
  year={2020},
  publisher={IEEE}
}

@inproceedings{yang2021learning,
  title={Learning dynamics via graph neural networks for human pose estimation and tracking},
  author={Yang, Yiding and Ren, Zhou and Li, Haoxiang and Zhou, Chunluan and Wang, Xinchao and Hua, Gang},
  booktitle={Proceedings of the IEEE/CVF conference on computer vision and pattern recognition},
  pages={8074--8084},
  year={2021}
}

@article{lan2021medial,
  title={Medial IPC: accelerated incremental potential contact with medial elastics},
  author={Lan, Lei and Yang, Yin and Kaufman, Danny and Yao, Junfeng and Li, Minchen and Jiang, Chenfanfu},
  journal={ACM Transactions on Graphics},
  volume={40},
  number={4},
  year={2021}
}

@inproceedings{dou2023tore,
  title={Tore: Token reduction for efficient human mesh recovery with transformer},
  author={Dou, Zhiyang and Wu, Qingxuan and Lin, Cheng and Cao, Zeyu and Wu, Qiangqiang and Wan, Weilin and Komura, Taku and Wang, Wenping},
  booktitle={Proceedings of the IEEE/CVF International Conference on Computer Vision},
  pages={15143--15155},
  year={2023}
}

@article{lan2020medial,
  title={Medial elastics: Efficient and collision-ready deformation via medial axis transform},
  author={Lan, Lei and Luo, Ran and Fratarcangeli, Marco and Xu, Weiwei and Wang, Huamin and Guo, Xiaohu and Yao, Junfeng and Yang, Yin},
  journal={ACM Transactions on Graphics (TOG)},
  volume={39},
  number={3},
  pages={1--17},
  year={2020},
  publisher={ACM New York, NY, USA}
}

@article{10.1145/3550454.3555453, 
author = {Xin, Shiqing and Wang, Pengfei and Xu, Rui and Yan, Dongming and Chen, Shuangmin and Wang, Wenping and Zhang, Caiming and Tu, Changhe}, 
title = {SurfaceVoronoi: Efficiently Computing Voronoi Diagrams Over Mesh Surfaces with Arbitrary Distance Solvers}, 
year = {2022}, 
issue_date = {December 2022}, 
publisher = {Association for Computing Machinery}, 
address = {New York, NY, USA}, 
volume = {41}, 
number = {6}, 
issn = {0730-0301}, 
url = {https://doi.org/10.1145/3550454.3555453}, 
doi = {10.1145/3550454.3555453}, 
journal = {ACM Trans. Graph.}, 
month = nov, 
articleno = {185}, 
numpages = {12}
}

@incollection{cgal:pt-t3-25b,
  author = {Cl{\'e}ment Jamin and Sylvain Pion and Monique Teillaud},
  title = {{3D} Triangulations},
  publisher = {{CGAL Editorial Board}},
  edition = {{6.1}},
  booktitle = {{CGAL} User and Reference Manual},
  url = {https://doc.cgal.org/6.1/Manual/packages.html#PkgTriangulation3},
  year = 2025
}

@misc{blender-mat-addon,
  title = {blender-mat-addon},
  author = {Song, Shibo and Wang, Ningna},
  howpublished = "\url{https://github.com/songshibo/blender-mat-addon}",
  year = {2023}
}

@article{thiery2013sphere,
  title={Sphere-meshes: Shape approximation using spherical quadric error metrics},
  author={Thiery, Jean-Marc and Guy, {\'E}milie and Boubekeur, Tamy},
  journal={ACM Transactions on Graphics (TOG)},
  volume={32},
  number={6},
  pages={1--12},
  year={2013},
  publisher={ACM New York, NY, USA}
}

@inproceedings{petrov2024gem3d,
  title={Gem3d: Generative medial abstractions for 3d shape synthesis},
  author={Petrov, Dmitry and Goyal, Pradyumn and Thamizharasan, Vikas and Kim, Vladimir and Gadelha, Matheus and Averkiou, Melinos and Chaudhuri, Siddhartha and Kalogerakis, Evangelos},
  booktitle={ACM SIGGRAPH 2024 Conference Papers},
  pages={1--11},
  year={2024}
}

@article{yang2020p2mat,
  title={P2MAT-NET: Learning medial axis transform from sparse point clouds},
  author={Yang, Baorong and Yao, Junfeng and Wang, Bin and Hu, Jianwei and Pan, Yiling and Pan, Tianxiang and Wang, Wenping and Guo, Xiaohu},
  journal={Computer Aided Geometric Design},
  volume={80},
  pages={101874},
  year={2020},
  publisher={Elsevier}
}

@article{yan2016erosion,
  title={Erosion thickness on medial axes of 3D shapes},
  author={Yan, Yajie and Sykes, Kyle and Chambers, Erin and Letscher, David and Ju, Tao},
  journal={ACM Transactions on Graphics (TOG)},
  volume={35},
  number={4},
  pages={1--12},
  year={2016},
  publisher={ACM New York, NY, USA}
}

@inproceedings{lin2021point2skeleton,
  title={Point2skeleton: Learning skeletal representations from point clouds},
  author={Lin, Cheng and Li, Changjian and Liu, Yuan and Chen, Nenglun and Choi, Yi-King and Wang, Wenping},
  booktitle={Proceedings of the IEEE/CVF conference on computer vision and pattern recognition},
  pages={4277--4286},
  year={2021}
}

@article{Liu1989OnTL,
  title={On the limited memory BFGS method for large scale optimization},
  author={Dong C. Liu and Jorge Nocedal},
  journal={Mathematical Programming},
  year={1989},
  volume={45},
  pages={503-528},
  url={https://api.semanticscholar.org/CorpusID:5681609}
}

@inproceedings{Dey1998TopologyPE,
  title={Topology preserving edge contraction},
  author={Tamal K. Dey and Herbert Edelsbrunner and Sumanta Guha and Dmitry V. Nekhayev},
  year={1998},
  url={https://api.semanticscholar.org/CorpusID:50333417}
}

@article{Hou2023DCUDF,
	author = {Hou, Fei and Chen, Xuhui and Wang, Wencheng and Qin, Hong and He, Ying},
	title = {Robust Zero Level-Set Extraction from Unsigned Distance Fields Based on Double Covering},
	year = {2023},
	publisher = {Association for Computing Machinery},
	address = {New York, NY, USA},
	volume = {42},
	number = {6},
	issn = {0730-0301},
	doi = {10.1145/3618314},
	journal = {ACM Trans. Graph.},
	month = {dec},
	articleno = {245},
	numpages = {15},
}

@ARTICLE{11079233,
  author={Chen, Xuhui and Yu, Fugang and Hou, Fei and Wang, Wencheng and Zhang, Zhebin and He, Ying},
  journal={IEEE Transactions on Visualization and Computer Graphics}, 
  title={DCUDF2: Improving Efficiency and Accuracy in Extracting Zero Level Sets From Unsigned Distance Fields}, 
  year={2025},
  volume={31},
  number={10},
  pages={9052-9065},
  doi={10.1109/TVCG.2025.3588659}}

@inproceedings{zhu2020deep,
    title={Deep Fashion3D: A Dataset and Benchmark for 3D Garment Reconstruction from Single Images}, 
    booktitle={Computer Vision -- ECCV 2020},
    author={Heming, Zhu and Yu, Cao and Hang, Jin and Weikai, Chen and Dong, Du and Zhangye, Wang and Shuguang, Cui and Xiaoguang, Han},
    year={2020},
    publisher={Springer International Publishing},
    pages={512--530},
    isbn={978-3-030-58452-8}
}

@inproceedings{10.1145/258734.258849,
author = {Garland, Michael and Heckbert, Paul S.},
title = {Surface simplification using quadric error metrics},
year = {1997},
isbn = {0897918967},
publisher = {ACM Press/Addison-Wesley Publishing Co.},
address = {USA},
url = {https://doi.org/10.1145/258734.258849},
doi = {10.1145/258734.258849},
booktitle = {Proceedings of the 24th Annual Conference on Computer Graphics and Interactive Techniques},
pages = {209–216},
numpages = {8},
series = {SIGGRAPH '97}
}

@article{https://doi.org/10.1111/j.1467-8659.2009.01521.x,
author = {Yan, Dong-Ming and Lévy, Bruno and Liu, Yang and Sun, Feng and Wang, Wenping},
title = {Isotropic Remeshing with Fast and Exact Computation of Restricted Voronoi Diagram},
journal = {Computer Graphics Forum},
volume = {28},
number = {5},
pages = {1445-1454},
doi = {https://doi.org/10.1111/j.1467-8659.2009.01521.x},
url = {https://onlinelibrary.wiley.com/doi/abs/10.1111/j.1467-8659.2009.01521.x},
eprint = {https://onlinelibrary.wiley.com/doi/pdf/10.1111/j.1467-8659.2009.01521.x},
year = {2009}
}

@InProceedings{10.1007/978-3-642-33573-0_21,
author="L{\'e}vy, Bruno
and Bonneel, Nicolas",
editor="Jiao, Xiangmin
and Weill, Jean-Christophe",
title="Variational Anisotropic Surface Meshing with Voronoi Parallel Linear Enumeration",
booktitle="Proceedings of the 21st International Meshing Roundtable",
year="2013",
publisher="Springer Berlin Heidelberg",
address="Berlin, Heidelberg",
pages="349--366",
isbn="978-3-642-33573-0"
}

@software{levy2026graphitethree,
  author  = {L{\'e}vy, Bruno},
  title   = {{GraphiteThree}: An Experimental {3D} Modeler},
  year    = {2026},
  version = {v3-1.9.9},
  url     = {https://github.com/BrunoLevy/GraphiteThree},
  note    = {Built around the Geogram library}
}

@ARTICLE{10845125,
  author={Wang, Pengfei and Song, Jiantao and Wang, Lei and Xin, Shiqing and Yan, Dong-Ming and Chen, Shuangmin and Tu, Changhe and Wang, Wenping},
  journal={IEEE Transactions on Visualization and Computer Graphics}, 
  title={Towards Voronoi Diagrams of Surface Patches}, 
  year={2025},
  volume={31},
  number={10},
  pages={6810-6823},
  doi={10.1109/TVCG.2025.3531445}}

@article{10.1145/3731166,
author = {Jiang, Jing-En and Wang, Hanxiao and Zhao, Mingyang and Yan, Dong-Ming and Chen, Shuangmin and Xin, Shiqing and Tu, Changhe and Wang, Wenping},
title = {DeFillet: Detection and Removal of Fillet Regions in Polygonal CAD Models},
year = {2025},
issue_date = {August 2025},
publisher = {Association for Computing Machinery},
address = {New York, NY, USA},
volume = {44},
number = {4},
issn = {0730-0301},
url = {https://doi.org/10.1145/3731166},
doi = {10.1145/3731166},
journal = {ACM Trans. Graph.},
month = jul,
articleno = {115},
numpages = {19}
}

\appendix
\appendix

\section{Gradient Derivation and Closed-Form Solution}
\label{app:gradient}

This appendix provides the explicit gradients required by the L-BFGS 
optimization in Section~\ref{sec:OptimalVertexPlacement}, as well as 
the closed-form solution available when no invaginated cells are 
involved. L-BFGS requires only first-order gradient information, so 
no explicit Hessian is derived.

\paragraph{Optimization variables.}
The optimization is performed over 
$\overline{v} = (v_x, v_y, v_z, r)^\top \in \mathbb{R}^4$, i.e., the 
sphere center $v$ and radius $r$. All surface-side quantities 
($s_i$, $p$, $\mathbf{n}$, and the associated region $\mathcal{A}$) 
are fixed constants inherited at the moment of edge collapse.

\paragraph{Facet-level decomposition.}
The surface region within each RVD cell is composed of planar triangle 
facets. By linearity of differentiation, the total gradient is the sum 
of per-facet gradients:
\begin{equation}
    \nabla_{\overline{v}} E_{\text{Fidelity}}(\overline{v}) 
    = \sum_{T \in \mathcal{A}} \nabla_{\overline{v}} E_T(\overline{v}),
\end{equation}
where $T$ denotes a triangle facet with area $|T|$. We treat regular 
and invaginated cells separately below.

\subsection{Regular Cells: Quadratic Form and Gradient}
\label{app:gradient:regular}

For a facet $T$ in a regular cell with unit normal $\mathbf{n}$, the 
per-facet energy reduces to a quadratic form in $\overline{v}$:
\begin{equation}
    E_T(\overline{v}) 
    = \overline{v}^\top \mathbf{A}_T \overline{v} 
    + \mathbf{b}_T^\top \overline{v} + c_T,
\end{equation}
with coefficients
\begin{equation}
    \mathbf{A}_T = |T| 
    \begin{pmatrix} 
        \mathbf{n}\mathbf{n}^\top & \mathbf{n} \\ 
        \mathbf{n}^\top & 1 
    \end{pmatrix}, \quad
    \mathbf{b}_T = -2 \, |T| \, (\mathbf{n}^\top \bar{p}_T)
    \begin{pmatrix} \mathbf{n} \\ 1 \end{pmatrix}, \quad
    c_T = \int_T (\mathbf{n}^\top p)^2 \, dA,
\end{equation}
where $\bar{p}_T$ is the centroid of $T$. The gradient is linear:
\begin{equation}
    \nabla_{\overline{v}} E_T(\overline{v}) 
    = 2 \mathbf{A}_T \overline{v} + \mathbf{b}_T.
    \label{eq:grad:regular}
\end{equation}
By additivity, contributions from all regular facets can be 
pre-aggregated once into global coefficients before optimization:
\begin{equation}
    \mathbf{A}_{\text{reg}} = \sum_{T \in \mathcal{A}_{\text{reg}}} 
    \mathbf{A}_T, \quad
    \mathbf{b}_{\text{reg}} = \sum_{T \in \mathcal{A}_{\text{reg}}} 
    \mathbf{b}_T.
\end{equation}

\subsection{Invaginated Cells: Nonlinear Gradient}
\label{app:gradient:invaginated}

For a facet $T$ in an invaginated cell with generating site $s_i$, the 
energy $E_T(\overline{v}) = |T| (\|v - s_i\| - r)^2$ is nonlinear. 
Computing its gradient:
\begin{equation}
    \frac{\partial E_T}{\partial v} 
    = 2 |T| (\|v - s_i\| - r) 
    \cdot \frac{v - s_i}{\|v - s_i\|}, 
    \qquad
    \frac{\partial E_T}{\partial r} 
    = 2 |T| (r - \|v - s_i\|).
\end{equation}
Stacked as a 4D gradient:
\begin{equation}
    \nabla_{\overline{v}} E_T(\overline{v}) 
    = 2 |T| (\|v - s_i\| - r)
    \begin{pmatrix} 
        (v - s_i) / \|v - s_i\| \\ 
        -1 
    \end{pmatrix}.
    \label{eq:grad:invaginated}
\end{equation}
This gradient depends on the current $\overline{v}$ through both the 
scalar factor and the direction, so it is recomputed at every L-BFGS 
iteration. For numerical stability, the denominator $\|v - s_i\|$ is 
clamped to $\max(\|v - s_i\|, \epsilon)$ with $\epsilon = 10^{-8}$.

\subsection{Topological Laplacian Smoothing Gradient}
\label{app:gradient:laplacian}

The Laplacian term is quadratic in $v$ and independent of $r$:
\begin{equation}
    \frac{\partial E_{\text{Lap}}}{\partial v} 
    = \frac{2}{|\mathcal{N}(v_1, v_2)|} 
    \sum_{u \in \mathcal{N}(v_1, v_2)} (v - u), \quad
    \frac{\partial E_{\text{Lap}}}{\partial r} = 0.
    \label{eq:grad:lap}
\end{equation}

\subsection{Closed-Form Solution in the Purely Quadratic Case}
\label{app:gradient:closedform}

When $\mathcal{A}$ contains no invaginated cells, the total energy is 
purely quadratic in $\overline{v}$ and can be solved in closed form 
without L-BFGS. Writing the Laplacian term in matrix form:
\begin{equation}
    E_{\text{Lap}}(\overline{v}) 
    = \overline{v}^\top \mathbf{A}_{\text{Lap}} \overline{v} 
    + \mathbf{b}_{\text{Lap}}^\top \overline{v} + c_{\text{Lap}},
\end{equation}
with
\begin{equation}
    \mathbf{A}_{\text{Lap}} = 
    \begin{pmatrix} \mathbf{I}_3 & \mathbf{0} \\ \mathbf{0}^\top & 0 \end{pmatrix}, 
    \quad
    \mathbf{b}_{\text{Lap}} = -\frac{2}{|\mathcal{N}(v_1, v_2)|} 
    \sum_{u \in \mathcal{N}(v_1, v_2)} 
    \begin{pmatrix} u \\ 0 \end{pmatrix},
\end{equation}
the total energy becomes
\begin{equation}
    E(\overline{v}) 
    = \overline{v}^\top \mathbf{A}_{\text{tot}} \overline{v} 
    + \mathbf{b}_{\text{tot}}^\top \overline{v} + \text{const},
\end{equation}
where
\begin{equation}
    \mathbf{A}_{\text{tot}} = \mathbf{A}_{\text{reg}} + \lambda \mathbf{A}_{\text{Lap}}, 
    \quad
    \mathbf{b}_{\text{tot}} = \mathbf{b}_{\text{reg}} + \lambda \mathbf{b}_{\text{Lap}}.
\end{equation}
Setting $\nabla_{\overline{v}} E = 2 \mathbf{A}_{\text{tot}} \overline{v} + \mathbf{b}_{\text{tot}} = \mathbf{0}$ 
yields the closed-form minimizer:
\begin{equation}
    \overline{v}^\ast 
    = -\frac{1}{2} \mathbf{A}_{\text{tot}}^{-1} \mathbf{b}_{\text{tot}}.
    \label{eq:closedform}
\end{equation}
Since $\mathbf{A}_{\text{tot}} \in \mathbb{R}^{4 \times 4}$, 
Eq.~\ref{eq:closedform} is obtained by solving a small $4 \times 4$ 
linear system and incurs negligible cost per edge collapse.

\subsection{Total Gradient in the Nonlinear Case}
\label{app:gradient:total}

When $\mathcal{A}$ contains invaginated cells, the total gradient used 
by L-BFGS is assembled as:
\begin{equation}
\begin{aligned}
    \nabla_{\overline{v}} E(\overline{v}) 
    = {} & \underbrace{2 \mathbf{A}_{\text{reg}} \overline{v} 
    + \mathbf{b}_{\text{reg}}}_{\text{regular (pre-aggregated)}} \\
    & + \underbrace{\sum_{T \in \mathcal{A}_{\text{inv}}} 
    2 |T| (\|v - s_{i(T)}\| - r)
    \begin{pmatrix} 
        (v - s_{i(T)}) / \|v - s_{i(T)}\| \\ 
        -1 
    \end{pmatrix}}_{\text{invaginated (per-iteration)}} \\
    & + \lambda 
    \begin{pmatrix} 
        \dfrac{2}{|\mathcal{N}(v_1, v_2)|} 
        \sum_{u \in \mathcal{N}(v_1, v_2)} (v - u) \\[2pt] 
        0 
    \end{pmatrix}.
\end{aligned}
\end{equation}
where $s_{i(T)}$ is the generating site of the invaginated cell 
containing $T$. The per-iteration cost is proportional to 
$|\mathcal{A}_{\text{inv}}| + |\mathcal{N}(v_1, v_2)|$.

\end{document}